\documentclass[apj]{emulateapj}

\usepackage{hyperref}
\usepackage{textcomp}
\usepackage{bm,color}
\usepackage{verbatim}
\usepackage{amsmath}
\usepackage{ulem}
\usepackage{subfigure}
\usepackage{multirow}
\usepackage{stackengine}

\newcommand{\secref}{Sec.}

\newcommand{\figref}{Figure}
\newcommand{\tabref}{Table}

\newcommand{\figrefs}{Figures}

\newcommand{\cmb}{CMB}	
\newcommand{\pla}{\textit{Planck}}

\newcommand{\ie}{i.e.}
\newcommand{\eg}{e.g.}
\newcommand\gpedit[1]{\normalfont{#1}}
\received{\today}
\revised{\today}
\accepted{\today}

\shorttitle{Forecasting Polarized Radio Sources for \cmb\   observations}
\shortauthors{Puglisi et al.}
 
\begin{document}
\normalem 
\title{Forecasting the Contribution of Polarized Extragalactic Radio Sources in  \cmb\ observations}

\author{
G. Puglisi\altaffilmark{1,2},
V. Galluzzi\altaffilmark{3,5},
L. Bonavera\altaffilmark{4},
J. Gonzalez-Nuevo\altaffilmark{4},
A. Lapi\altaffilmark{1},
M. Massardi\altaffilmark{3,6},
F. Perrotta \altaffilmark{1},
C. Baccigalupi\altaffilmark{1,2},
A. Celotti\altaffilmark{1,2,7},
L. Danese\altaffilmark{1}
}

\affiliation{$^1$SISSA- International School for Advanced Studies, Via Bonomea 265, 34136 Trieste, Italy}
\affiliation{$^2$INFN-National Institute for Nuclear Physics, Via Valerio 2, I-34127 Trieste, Italy}
\affiliation{$^3$INAF, Istituto di Radioastronomia, Via Piero Gobetti 101, I-40129 Bologna, Italy}
\affiliation{$^4$Departamento de F\'isica, Universidad de Oviedo, C. Federico Garc\'ia Lorca 18, 33007 Oviedo, Spain}
\affiliation{$^5$Dipartimento di Fisica e Astronomia, Universit\`a di Bologna, via Gobetti 93/2, I-40129 Bologna, Italy}
\affiliation{$^6$Italian Alma Regional Centre, Istituto di Radioastronomia, Via Piero Gobetti 101, I-40129 Bologna, Italy }
\affiliation{$^7$INAF, Osservatorio Astronomico di Brera, via Bianchi 46, 23807 Merate (LC)}
\begin{abstract}
We combine the latest datasets obtained with different surveys to study the frequency dependence of polarized emission coming from Extragalactic Radio Sources (ERS). We consider  data over a very wide frequency range starting from $1.4$ GHz up to $217$ GHz. This range is particularly interesting since it overlaps the frequencies of the current and forthcoming Cosmic Microwave Background (\cmb) experiments. Current data suggest that at high radio frequencies, ($ \nu \geq 20$ GHz) the fractional polarization of ERS does not depend on the total flux density. Conversely, \gpedit{recent datasets indicate a moderate increase of polarization fraction as a function of frequency, physically motivated by the fact that  Faraday depolarization is expected to be less relevant at high radio-frequencies.}
We compute ERS number counts using updated models based on recent data, and we forecast the contribution of unresolved ERS in CMB polarization spectra.
Given the expected sensitivities and the observational patch sizes of forthcoming \cmb\ experiments about $\sim 200 $ ( up to $\sim 2000  $ ) polarized ERS are expected to be detected. Finally, we assess that polarized ERS can contaminate the cosmological B-mode polarization if the tensor-to-scalar ratio is $< 0.05$ and they have to be robustly  controlled  to \emph{de-lens} \cmb\ B-modes at the arcminute angular scales.
\end{abstract}

\thanks{
	\footnotesize{
		Corresponding authors: Giuseppe~Puglisi, giuspugl@sissa.it
		 \email{giuspugl@sissa.it}
	}
}



\section{Introduction}
\setcounter{footnote}{0}

\gpedit{The Cosmic Microwave Background (CMB) is a relic radiation generated at the decoupling of matter and radiation as the temperature of the Universe dropped below ${3000}$ K. Its temperature and polarization anisotropies  can be exploited to probe the early stages of the Universe when an exponential expansion, the so called \emph{inflation} might have occurred \citep{1981PhRvD..23..347G, 1982PhLB..117..175S}.} 

\gpedit{Since last decades, several experiments have tried to measure the \cmb\ polarized signal  in order to find the imprints on its polarized  anisotropies of a stochastic background of primordial gravitational waves (PGW) that might have been produced during the inflationary phase. Polarization anisotropies are commonly decomposed into two scalar quantities called \emph{E-} and \emph{B-modes} \citep{1997PhRvL..78.2054S,Hu:1997hv}, and  to date, lots of efforts have been made to observe the latter since their amplitude at degree scale is expected to come mainly from PGW. }

\gpedit{On one hand, E-mode photons get deflected via gravitational interaction by intervening matter of large scale structures during the path toward us, producing the so called  \emph{lensing B-modes} at arcminute scale. Lensing  B-modes  have been observed since four years  \citep{2014ApJ...794..171T,2016arXiv161002360L,sptpol2015,ThePOLARBEARCollaboration2017} with better and better accuracy and they represent a powerful tool to probe the large scale structure of our Universe. On the contrary, the primordial B-mode amplitude is unknown and  is quantified by the \emph{tensor-to-scalar ratio}, $r$, that relates the amplitude of tensor perturbations of the space time metric, \eg\ PGW, with respect to the scalar perturbations.  The joint collaboration of BICEP2 and \pla\  yielded so far the latest upper limit on $r< 0.07$ at $95\%$ confidence level \citep{PhysRevLett.114.101301}. Meaning that the primordial B-mode amplitude could  be even lower than the lensing one.}

\noindent \gpedit{ To date, several challenges have prevented to detect primordial B-modes mostly because of the diffuse polarized radiation coming from the Milky Way, known as \emph{Galactic Foregrounds}.  The list of  Galactic foregrounds is long and includes anything emitting at sub-millimeter  wavelengths between us and the \cmb: thermal dust, synchrotron radiation, free-free and several molecular line  emissions \citep{pldiffuse2015}. 
All these emissions are partially polarized and the main contribution comes from synchrotron and dust \citep[both polarized up to   $ 20\% $ level][]{2016A&A...586A.133P,pldiffuselow}. 
At high-frequency ($\nu > 90 $ GHz), such a large polarization degree is produced by thermal dust grains aligning along the Galactic magnetic field lines. At low frequencies ($\nu \lesssim 70 $ GHz), cosmic electrons spiralling into the Galactic magnetic field produce synchrotron radiation. 
  Molecular lines are expected to be polarized at lower levels  $\lesssim 1 \%$ \citep{Goldreich1981,2017MNRAS.469.2982P}, whereas free-free emission can be essentially considered unpolarized. This is the justification of the recent efforts aimed at observing the \cmb\ polarization in a very wide range of frequencies and at accurately characterizing both the spatial and frequency distribution of each Galactic polarized foreground. Moreover, such an investigation allows to design algorithms known as \emph{component separation} or \emph{foreground cleaning} techniques to extract B-modes  out of a multi-frequency experimental setup.}
 
For these reasons, (i) more focal plane pixels in multiple telescopes  are needed to increase sensitivity and  (ii) multi-band polarization measurements are required to recover the cosmic signal from the Galactic one via {component separation}. \gpedit{As  the focal plane will encode larger and larger number of detectors, the next   \emph{stages} in \cmb\ experiment sensitivity will be achieved by more accurately measuring $r$. To date, several ground based experiments are updating their focal planes to a step forward from the so called \cmb-Stage 2 (\cmb-S2) to  Stage 3 \citep[\cmb-S3][]{arnold2014,advactpol,spt3g}, including  up to $10,000$ detectors  observing up to $7 \%$ of the sky. The ultimate step for a B-mode detection from the ground is represented by \cmb\-Stage 4 experiments \citep[\cmb-S4][]{2016arXiv161002743A}, which will account for up to $100,000$ detectors, observing half of the sky. \cmb-S4  aims at measuring $r$ with the target accuracy $\sigma(r)\sim 0.0005$ . } %

\gpedit{At smaller scales the Extragalactic Radio Sources (ERS) and star-forming dusty galaxies are the major contaminants \citep{Tucci2011}, although the latter can also largely contribute to large angular scales due to clustering  \citep{2015JCAP...06..018D}. }
 In this work, we mostly focus on the polarized emission of ERS. 
 To date, a few studies have been conducted regarding polarization of ERS at the frequencies of \cmb\ experiments (see \citet{2016IJMPD..2540005G} or \citet{Bonavera2017})  since polarization observations in the mm wavelength bands are more challenging than at cm bands (at $1.4\div 20$  GHz) and extrapolations are very common in this field of research \citep{Tucci2012}.  

\noindent The mechanism behind the polarized emission of radio sources is mostly due to synchrotron radiation sourced by an Active Galactic Nucleus (AGN), where a central super-massive black hole ($10^6\div 10^9 M_{\odot}$) is hosted. Most of the energy of an AGN comes from the gravitational potential energy of the material located in a thin surrounding accretion disc, released as the matter falls into the central black hole. Another component is constituted by  jets (usually paired) of material ejected toward the polar directions from the black hole. Jets are observed to be very collimated and can travel  very large distances. Therefore, radio-galaxies seldom present double structures referred as \emph{lobes} constantly fed by the jets of new energetic particles and magnetic energy. 

\noindent Depending on which components dominates the emission, such complex objects can obviously appear with different morphologies and therefore be grouped in different observational categories.
One of the most important distinction is related to the different orientations an AGN can be observed with respect to the line of sight (see \citet{DeZotti2010} for a wide review). If edge-on, the torus obscures the core and the inner disc, so that the emission is dominated by the optically thin radio lobes presenting a steep spectral index  $\alpha$ at low frequencies $1\div 5$ GHz\footnote{The radio-source flux is described by a power law $S_{\nu} \propto \nu^{-\alpha }$, and the threshold between flat and steep spectral behaviour is commonly fixed at $\alpha = 0.5$.}. Objects with $\alpha > 0.5$ are commonly referred as \emph{Steep  Spectrum Radio Quasars} (SSRQs) and, generally, their optical counterpart is an elliptical galaxy. If seen pole-on, the brightness is dominated by the approaching jet, the emission looks compact and it is mostly Doppler boosted since particles move at relativistic speeds. The emission is optically thick, does not contain many optical features in the continuum but is characterized by  a \emph{flat } spectrum ($\alpha < 0.5$). Similar sources are called \emph{Flat Spectrum Radio Quasars} (FSRQs). 

\noindent However, each source presents both the components, \ie\ a flat-spectrum core and extended steep-spectrum lobes, and it can be easily understood that a simple-power law cannot be applied to resemble the  large radio frequency range \citep{Massardi2011,Bonaldi2013}. External and self-absorption, from free-free and synchrotron, may affect and change the dependence of $S_{\nu}$, so that the spectrum could increase as a function of frequency \citep{Galluzzi2017a}.

\noindent  There is an increasing interest on polarization of ERS at high-radio frequencies not only to better understand the physics behind the emitting system, \eg\ the degree of ordering of the magnetic field, the direction of its field lines \citep{Tucci2011}, but also because polarized ERS will be largely detected by forthcoming \cmb\ experiments. Furthermore, the ERS contaminating signal in the polarization power spectra cannot be neglected to assess the power spectrum of lensing B-modes. This is the reason why recent works in the literature can be found addressing this issue: \citet{2015JCAP...06..018D,2016arXiv160907263D} predicted the contribution in polarization both for ERS and dusty galaxies at frequency channels of the Cosmic ORigin Explorer (CORE) satellite; \citet{2013MNRAS.432..728C} estimated for future \cmb\ missions the contamination produced by radio and far-Infrared sources at the level of bispectrum considering different shapes of the primordial non-Gaussianity  parameter, $f_{nl}$.

In \autoref{sec:data} we describe the datasets we combine in order to determine the polarization dependence as a function of frequency, discussed in \autoref{sec:polfrac}. 
In \autoref{sec:numbcounts}, we present the models  for number counts adopted in this analysis. In \autoref{sec:4casts}, we show the results of a forecast package we developed  to assess the contamination of polarized ERS in terms of \cmb\ power spectra given the nominal specifics  of current and forthcoming \cmb\ experiments.  Finally, we devote \autoref{sec:conclusions} to discuss and summarize  our results.

\section{Data}\label{sec:data} 
\noindent In this section, we present the data collected from publicly available  catalogues. The data, summarized in \tabref\ \ref{tab:catalogues}, have been used to characterize the polarization fraction of ERS in about two orders of magnitude in frequency range (\ie\  from $1.4 $ to $217$ GHz). 

\begin{table*}[htb]
\centering
{\footnotesize
\caption{Summary of the catalogues that we use  \secref\ref{sec:4casts}. }\label{tab:catalogues}
\begin{tabular}{lcccccc}
		\hline
      
 	& Frequency [GHz]& Sky Region			& FWHM & Detect. flux & $90 \%$ Compl.& $\#$ Sources \\
\hline
NVSS& 1.4 			 & $\delta >-40 ^{\circ}$&	$45''$ &$0.29$ mJy$/\mathrm{beam}$ &$2.3$ mJy & $1.8 \times 10^6$\\
S-PASS &2.3 & $\delta <-1 ^{\circ}$			 & $8.9 '$ &$1$ mJy$/\mathrm{beam}$ & $420$ mJy &$533$ \\
JVAS& 8.4&$\delta\geq 0^{\circ},\, |b|\geq2.5^{\circ} $&$0.2{''}$ &$50$ mJy &$200$ mJy   & 2720 \\
CLASS &8.4 & $0\geq\delta\geq 70^{\circ}$ & $0.2{''}$& $20$ mJy& $30$ mJy&  16503 \\
AT20G & $4.8$, $8.6$, $20$& $\delta<0^{\circ},\, |b|<1.5^{\circ}$ & $10{''},6{''} , 11{''} $& $40 $ mJy& $100$  mJy$/\mathrm{beam}$ & 5890\\
\multirow{2}{*}{VLA }& $4.8$, $8.5$,&$ \delta >-15^{\circ}$& $12''$, $6''$,&  $0.7$, $0.3$, & $40$ mJy &$159$ \\
& $22.5$, $43.5$& &  $4''$,$2''$&  $0.9$, $1.2$ mJy$/\mathrm{beam}$ & &  \\
PACO & $20$ & Ecl. lat.$<-65^{\circ}$&$11{'' }$ & $40 $ mJy&$200$ mJy& $104 $ \\
XPOL-IRAM &$86$& $\delta > 30^{\circ}$ & $28''$ & $0.5$ Jy &$1$ Jy & 145 \\
\multirow{3}{*}{PCCS2 }& $30$, $44$, & &$32.4'$, $27.1'$,& $117$,$229$,& $427$,$692$,& $1560$,$934$, \\
					  &$70$, $100$, & Full sky  &  $13.3'$, $9.7'$, & $225$, $106$,    & $501$,$269$, & $1296$,$1742$,\\
					  & $143$, $217$&   &    $7.3'$, $5.0'$ &   $75$,$81$ mJy & $177$,$152$ mJy& $2160$,$2135$  \\
\hline
\end{tabular}
}
\end{table*}

\subsection{The S-PASS/NVSS joint catalogue}
\noindent The S-band Polarization All Sky Survey (S-PASS) survey observed the Southern sky with declination $\delta <-1^{\circ} $ at 2.3 GHz with full width at half maximum (FWHM) of 8.9 arcmin both in total intensity and polarization  using  the $64 $ m  Parkes Radio Telescope. \citet{Lamee2016} cross-matched it with the NRAO/VLA Sky Survey, \citet[NVSS][]{Condon1998}, at $1.4 $GHz (45 arcsec (FWHM) and rms total brightness fluctuations of $\sim 0.29\, \mathrm{m Jy \,beam^{-1} }$). \citet{Lamee2016} aimed at generating a novel and independent polarization catalogue\footnote{\url{http://vizier.cfa.harvard.edu/viz-bin/VizieR?-source=J/ApJ/829/5}} enclosing 533 bright ERS \gpedit{ at $2.3$ GHz with polarized flux-density stronger than $420\,\mathrm{m Jy }$}. 

\subsection{The JVAS/CLASS 8.4 GHz catalogue}
\noindent  We used the data from the JVAS/CLASS 8.4-GHz catalogue \citet{Jackson2007}\footnote{\url{http://vizier.cfa.harvard.edu/viz-bin/VizieR?-source=J/MNRAS/376/371}}, which combined data taken from the Jodrell-VLA Astrometric Survey (JVAS) and the Cosmic Lens All-Sky Survey (CLASS) both observing at $8.4$ GHz. \gpedit{The former detected 2720 sources stronger than $200$ mJy in total intensity at $5$ GHz and $\delta \geq 0^{\circ}$, masking the Galactic mid-plane at Galactic latitude $|b|\geq2.5 ^{\circ}$. To complement JVAS, CLASS consisted of all sources with a fainter $5$ GHz flux, \ie\ $S>30$ mJy observed in a sky region between $0^{\circ} \leq \delta \leq 70^{\circ}$. Combining the two surveys, a sample of  $16\,503$ FSRQ intensity fluxes  has been collected.}

\gpedit{\citet{2010MNRAS.401.1388J} were able to assess polarized fluxes only for a few objects from the 133 sources observed by WMAP at $22$ and $43$ GHz \citet[$S > 1$ Jy][]{2009ApJS..180..283W} with counterpart in the JVAS/CLASS catalogues. For the purposes of our work this sample was not large enough to be included in the following analysis.}

\gpedit{However, we exploit the data selection described by \citet{Pelgrims2015} that considered all the sources with polarized flux $\geq 1 $ mJy in order to obtain an unbiased sample of 3858 NED} identified sources. We selected 2829 sources classified by \citet{Pelgrims2015} as \emph{QSOs} and \emph{Radio Sources}. For a complete description of the catalogue and the surveys refer to  \citet{Jackson2007}.

\subsection{The AT20G Survey}
\noindent The Australia Telescope 20 GHz (AT20G) Survey observed blindly the Southern sky ($\delta < 0 ^{\circ}$ excluding the Galactic plane strip at $|b|< {1.5} ^{\circ}$) at 20 GHz with the Australia Telescope Compact Array (ATCA) from 2004 to 2009, \citep{Murphy2010}. The detected sources were followed up almost simultaneously at 4.8 and 8.6 GHz. The AT20G source catalogue\footnote{\url{http://vizier.cfa.harvard.edu/viz-bin/VizieR?-source=J/MNRAS/402/2403}} includes 5890 sources at 20 GHz above the total intensity detection limit of $40$ mJy, of which 3332 were detected at all the observing frequencies. \gpedit{Averaged on the whole area of the survey, the catalogue is $91 \% $ complete above } $100 \, \mathrm{mJy\, beam^{-1}} $ \citep{Murphy2010}. \gpedit{Polarization of sources was considered detected if the following criteria were satisfied: polarized flux density $P> 6 $ mJy or  at least three times larger than its rms error, and polarized fraction above $1 $ per cent. \citet{Massardi2011} presented an analysis  to characterize the radio spectral properties of the whole sample both  in total intensity and polarization, involving  $768$ sources detected at $20$ GHz ($467$ of them were also detected in polarization at $4.8$ and/or at $8.6$ GHz).} Given the goal of this work, we include polarized flux densities from $3332$ sources, $2444$ of them presenting a flat spectrum in total intensity, $|\alpha_5^8|<0.5$ and  the remaining $888$ a steep-spectrum sources ($|\alpha_5^8|>0.5$).

\subsection{The VLA observations}
\noindent \citet{Sajina2011} presented measurements\footnote{\url{http://vizier.u-strasbg.fr/viz-bin/VizieR?-source=J/ApJ/732/45}} in flux densities and polarization of 159 ERSs detected with the Very Large Array (VLA) at four frequency channels, $4.86,\, 8.46,\, 22.46,\,43.34$ GHz. This sample was selected from the AT20G one \citep{Murphy2010,Massardi2011} by requiring a flux density  $S> 40$ mJy in the equatorial field of the Atacama Cosmology Telescope (ACT) survey on a region at declination north of  $-15^{\circ} $ and excluding the Galactic plane. \gpedit{The aim of this program was firstly to characterize the spectra and variability both in total intensity and polarization of high-frequency-selected radio sources and to improve the estimation of the ERS contamination at high-frequency for \cmb\ experiments.}

\noindent \gpedit{ In $40\%$ of the whole sample they detected polarized flux density in all the bands, and observed an increasing trend of polarization fraction as a function of frequency, more evident for SSRQs. }

\subsection{PACO with ATCA and ALMA}
\noindent The Planck-ATCA Coeval Observations (PACO) project detected  464 sources selected from the AT20G catalogue during 65 epochs between July 2009 and August 2010, at frequencies ranging from 5.5 to 39 GHz with the ATCA. The sources were simultaneously observed (within 10 days) by the Planck satellite \citep{2011MNRAS.415.1597M,2011MNRAS.416..559B}. The project aimed at characterizing, together with Planck data, the variability and spectral behavior of sources over a wide frequency range (up to 857 GHz for some sources), in total intensity only. 
The catalogue includes a complete sample of 159 sources selected to be brighter than $200$ mJy at $\delta < 30 ^{\circ} $ (excluding the Galactic midplane $|b|< 5^{\circ}$). A sub-sample of 104 of these sources with ecliptic latitude $<-65 ^{\circ}$ (which coincides to one of the \emph{deep } patches most frequently scanned by the \pla\ satellite scanning strategy) has been re-observed with high sensitivity in polarization with ATCA in 2014 and 2016 in the $1.1 - 39$ GHz frequency range \citep{Galluzzi2017a}. 32 of them have been also followed up at 95 GHz onto 3 circular regions ($10 ^{\circ}$ of diameter) at ecliptic latitude $<-75 ^{\circ}$ with the Atacama Large Millimeter Array (ALMA)   to better characterize the polarization properties of ERS at the frequencies of many \cmb\ experiments and allowing an accurate study of few reference targets which could be exploited for calibration and validation of cosmological results.  Further details will be described in a companion paper (Galluzzi et al. 2018, in prep.). Data from both 20 and 95 GHz have been included in this analysis.

\subsection{First 3.5 mm Polarimetric Survey }
\noindent \citet{Agudo2010} presented for the first time  polarimetric data at $86$ GHz  of a sample of 145 flat spectrum radio galaxies at  different epochs (from 2005 July to 2009 October)\footnote{\url{http://vizier.u-strasbg.fr/viz-bin/VizieR?-source=J/ApJS/189/1}.}. The measurements have been performed by means of the XPOL polarimeter of the IRAM 30 m telescope, by selecting the sources observed from 1978 to 1994 \gpedit{at $\delta > 30 ^{\circ}$ whose total intensity was above $\gtrsim 1$ Jy. They detected above $>3 \sigma $ level  $1.5 \%$ linear and  $0.3 \%$  circular polarization degree respectively  for  $76 \% $ and $6 \%$ of the whole sample. Remarkably, they found a factor of $\sim 2$ excess in the polarization fraction at 86 GHz with respect to that measured at 15 GHz. }

\subsection{The Second Planck Catalogue of Compact Sources }

\noindent We  exploit data from latest Planck Catalogue of Compact Sources \citep[PCCS2][]{PlanckPCCS2}\footnote{\url{http://pla.esac.esa.int/pla/}} including polarimetric detection of sources between $30$ and $353$ GHz from August 2009 to August 2013. The total intensity $90 \% $ completeness ranges from $177$ to $692$ mJy in this regime of frequencies, allowing to detect thousands of sources matched both internally (between neighbor \pla\ channels) and with external catalogues.
\gpedit{On the contrary, the instrumental noise in polarization and the presence of polarized Galactic foregrounds limited the number of polarized sources to a few tens (with the exception of the $30$ GHz channel where 113 polarized sources were detected).}
 
It is straightforward to state that only sources with high fractional polarization have been detected by \pla\ and thus the statistics of ERS polarization can be biased upward. \citet{Bonavera2017} recently proposed a methodology to cope with this issue by means of applying a \emph{stacking technique} to  \pla\ data. \gpedit{They used as main sample the 30 GHz catalogue, consisting of 1560 sources above $S>427$ mJy at $90 \%$ completeness level and then followed the sample at higher \pla\ frequency maps. They further distinguished sources inside and outside the Galactic plane defined by the Planck Galactic mask \texttt{GAL060} $(f_{sky}\approx  60\%)$ and the exclusion of the Small and Large Magellanic clouds. } This technique has been already applied by \citet{Stil2014} to NVSS dataset to study the faint polarized signal of ERS detected in total intensity: the signal from many weak sources is co-added to achieve a statistical detection. \citet{Bonavera2017}  found that the ERS polarization fraction is approximately constant with frequency over the \pla\ frequency range. An alternative approach that attempts to overcome some of the intrinsic statistical limitations of the stacking technique have been recently exploited by \citet{2017arXiv171208412T} obtaining results comparable both with   \citet{Bonavera2017,2017MNRAS.472..628B} and with other ground based observations.  
 
 \noindent \gpedit{We used both data coming from the PCCS2 catalogue and from \citet{Bonavera2017}.}

\section{Model for Number Counts} \label{sec:numbcounts}

\begin{figure}
\includegraphics[scale=.44, trim=.3cm .3cm 0 0, clip=true]{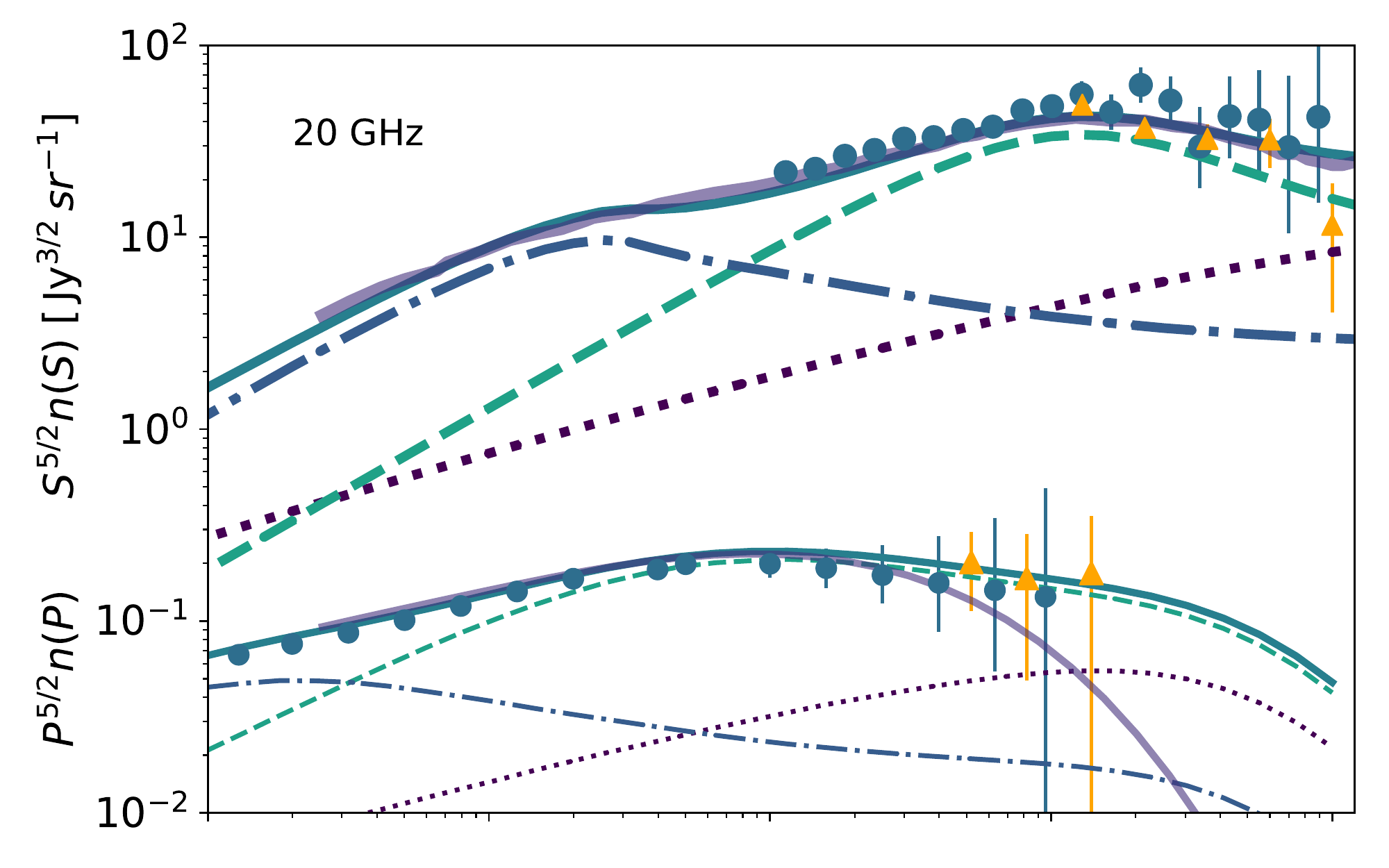}\\
\includegraphics[scale=.47, trim=0 0 0 1cm , clip=true]{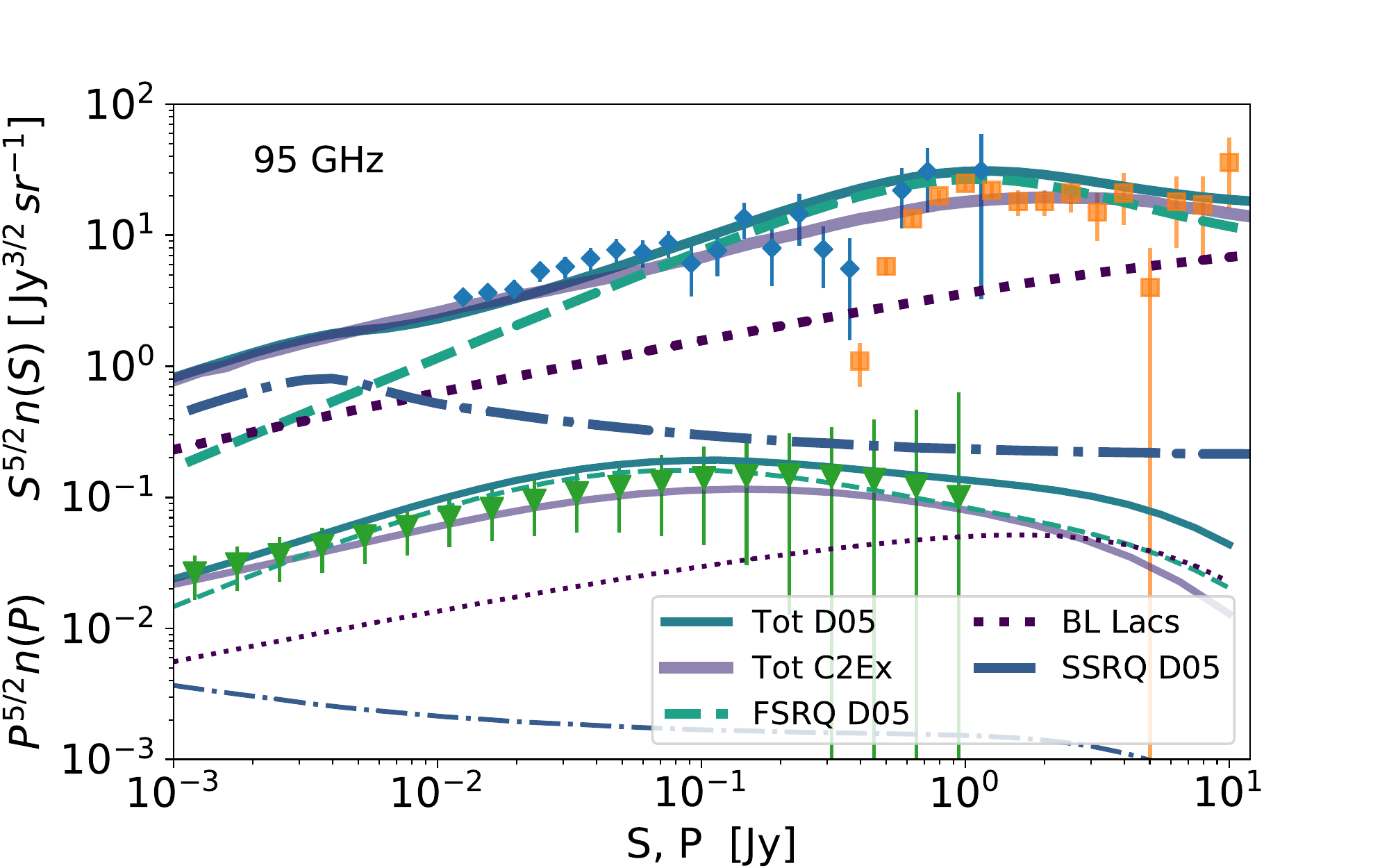}
\caption{Euclidean differential number counts at (top) 20 and (bottom) 95 GHz. Thick dotted, dashed, dot-dashed and solid lines are respectively the number counts of BL Lacs, FSRQs, SSRQs and their total contribution predicted by the D05 model \citep{2005A&A...431..893D}. The thick solid gray line show the number counts prediction from the  C2Ex model \citep{Tucci2011}. Thinner lines follow the same color scheme as the thick ones and refer to polarization number counts, computed via a convolution with a log-normal distribution function fitted from the data. Number counts estimates from several surveys are also shown. (top) The circle data point in the upper curves are data from AT20G \citep{Massardi2008}, whereas upper triangles are from WMAP5-yr survey  \citep[K-band ][]{2009MNRAS.392..733M}; in lower curves polarization number counts from a resampling of PACO data \citep[][, circles]{Galluzzi2017b} and from WMAP polarization point source catalogue \citep[][, upper triangles]{Lopez-Caniego2009}. (bottom panel) Diamonds in upper curves are number counts from SPT \citep{2013ApJ...779...61M}, squares are from \pla\ ERCSC catalogue \citep{PlanckearlyXII}; the lower triangles have been obtained from a bootstrap resampling of 32 polarized fluxes detected with PACO at 95 GHz.}
\label{fig:polcounts}
\end{figure}

 \gpedit{We adopted the evolutionary model proposed by \citet[][hereafter, D05]{2005A&A...431..893D} that describes the population properties of ERSs and dusty galaxies above $\nu \gtrsim 5 $ GHz. The model assumes a simple analytic luminosity evolution in order to fit the available data on local luminosity functions (LF), source counts\footnote{Available online \url{http://w1.ira.inaf.it/rstools/srccnt/srccnt_tables.html}. } and redshift distributions for sources down to few mJy. It determines the epoch-dependent LF starting from local LFs for several source populations. For each population the model adopts a different evolution laws estimating a set of free parameters from available data.  Recently, \citet{Bonato2017} and \citet{2017ApJ...842...95M} improved the predictions of D05 model by updating the LF and redshift evolution with state of art data of radio-emitting star-forming galaxies and AGNs.}   

The D05 model assumes a power-law spectrum for each considered population of ERS and each one is described by one (or at most two) constant spectral index. This simple assumptions could not hold anymore when large frequency ranges are taken into account. Departures from single power law spectra are expected because of (i) electron ageing (ii) transition from optically thick to optically thin regime, (iii) different components yielding different spectral contributions at different frequencies. Therefore, this \emph{simplified} model requires adjusting when source counts measurement are observed at frequencies $>40$ GHz.   
  
\citet{Tucci2011} showed that radio spectra in AGN cores can differ from a single power-law when large frequency intervals are considered.  In particular, they focused on the blazar spectra for which a steepening of the spectral index from $0.5$ to $1.2$ has been observed \citep{PlanckearlyXII,Aatrokoski2011} due to the transition from optically-thick to optically-thin synchrotron emission of AGN jets \citep{1966ApJ...146..621K,1979ApL....20...15B}. 
\gpedit{Therefore, \citet{Tucci2011} proposed the so called \emph{C2Ex} model that assumes a spectral break and different parameters for BL Lacs and FSRQs and allows } to properly fit the number counts especially at high-frequency ($\nu\geq 100$ GHz). Furthermore, \citet[][XXVI]{PlanckPCCS2} found that all radio sources observed at the Low Frequency Instrument (LFI) channels present flat and narrow spectral index distribution with $\alpha_{LFI}\lesssim 0.2$, whereas sources in the High Frequency Instrument (HFI) catalogues have a broader distribution showing a steeper spectral index, $\alpha_{HFI} \gtrsim 0.5 $ and these findings supports the scenario of BL Lac transition happening at larger frequencies  $\nu >100$ GHz with respect to the FSRQ one (at $10<\nu<100$ GHz). 
 
 \gpedit{We plot in \figref\ \ref{fig:polcounts} as thicker curves. } the differential number counts, $n(S)$,  predicted with D05 and C2Ex models respectively as blue and grey thick solid lines. The top (bottom) panel refers to number counts at $20\, (95)$ GHz\footnote{Source number counts for a wider range of frequencies are shown in \figref\ \ref{fig:counts}.}. We also plot the contributions estimated by the D05 model for BL Lacs, FSRQs, SSRQs  respectively as  dotted, dashed, dot-dashed lines. To compare the quantities with those expected in a Euclidean Universe, counts are normalized by a factor of $S^{5/2}$.
\gpedit{ The data points shown are number counts as measured by AT20G survey \citep[][blue circles]{Massardi2008},  from  South Pole Telescope \citep[SPT][blue diamonds]{Vieira2010,2013ApJ...779...61M}, from the Wilkinson Microwave Anisotropy Probe \citep[WMAP][yellow upper triangles]{2009MNRAS.392..733M} and from \pla\ \citep[][yellow squares]{PlanckearlyXII,PlanckIntermVII}.   }
 
 \gpedit{The lower thinner curves in \figref\  \ref{fig:polcounts} are Euclidean normalized differential polarized emission number counts, $P^{5/2} n(P)$, computed from polarized flux-density measurements and will be discussed in \secref\ \ref{sec:polfrac}. }
 
 \gpedit{By comparing the predictions from the two models, we find that both are in a reasonable agreement, with differences well below the uncertainties at $20$ GHz. However, as discussed above and shown in the bottom  panel of \figref\ \ref{fig:polcounts}, number counts estimated with D05 are systematically a factor of $\sim 2$ higher than the C2Ex ones at larger fluxes $100$ mJy, consistently with the findings of \citet{PlanckearlyXII}.} 

In the following, we make use of both D05 and C2Ex  models to assess respectively \emph{conservative} and \emph{realistic} estimates of polarized ERS to \cmb\ measurements. 

\section{Statistical properties of ERS polarization fraction} 
\label{sec:polfrac}
\begin{figure*}[htb]
\includegraphics[width=1\columnwidth]{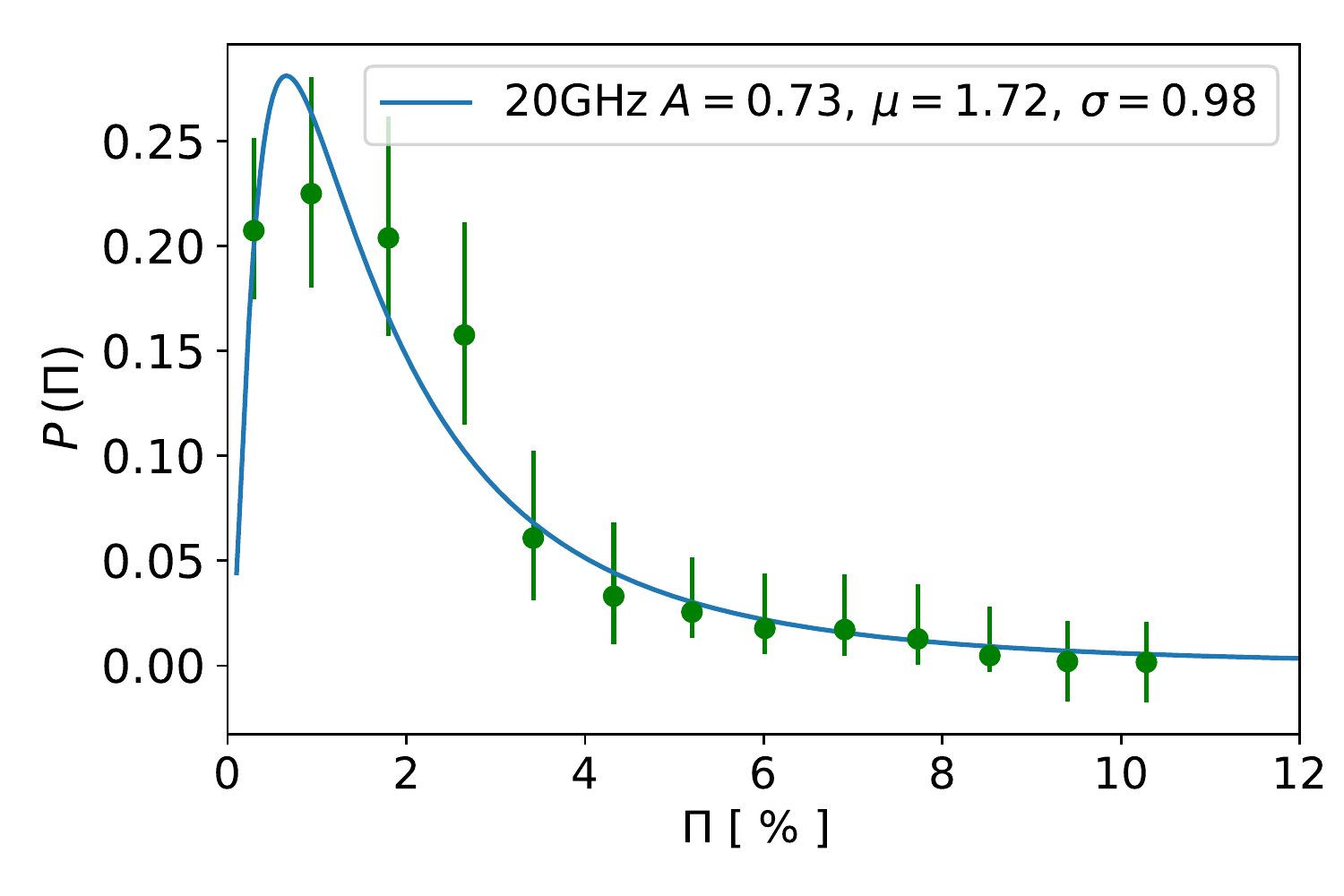}
\includegraphics[width=1\columnwidth]{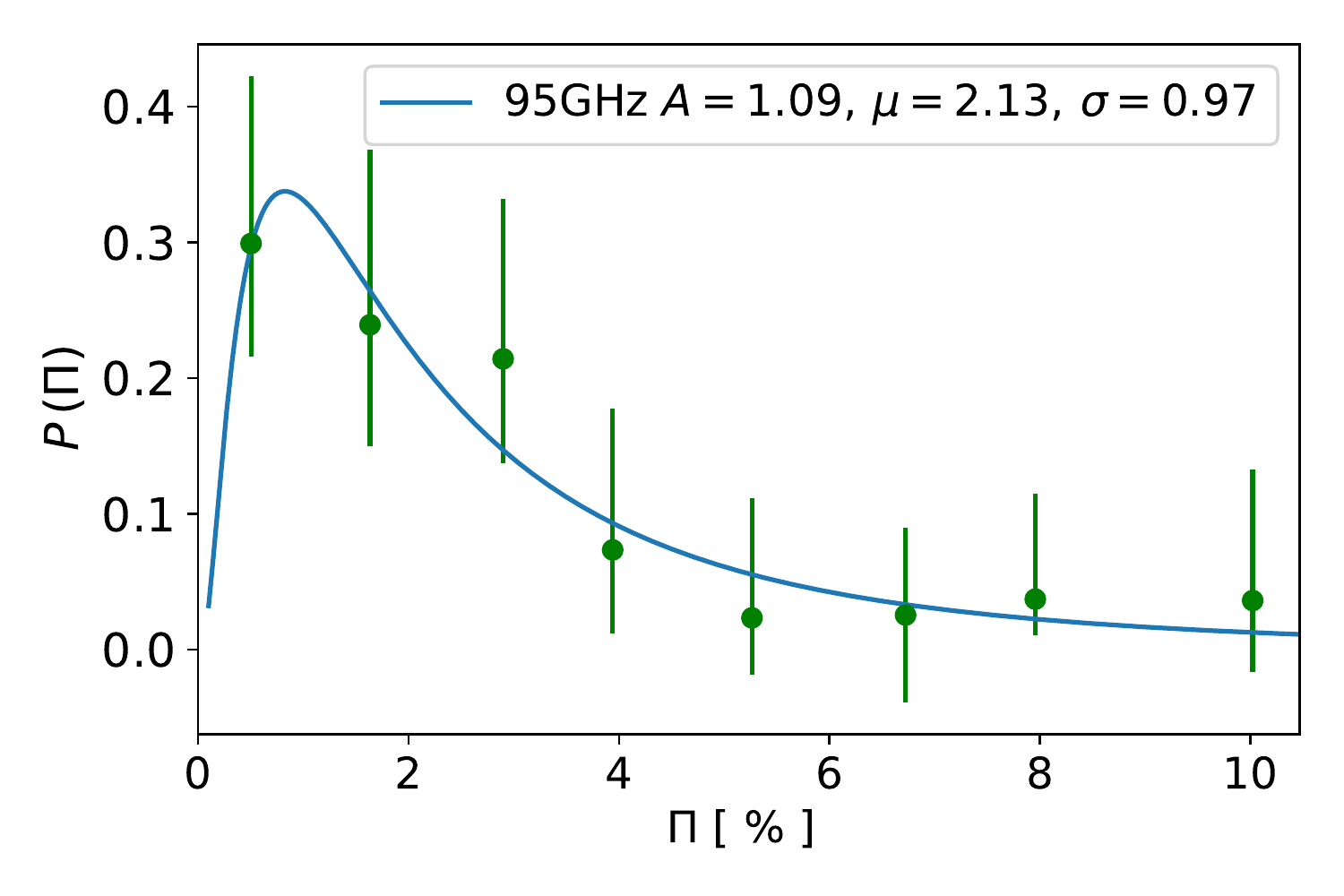}
\caption{ Distribution function of polarization fraction for data at 20 GHz (left) and at 95 GHz (right). The best fit values of log-normal parameters are shown. The reduced $\tilde{\chi}^2$ estimated from the fit is 0.13 and 0.15 respectively for left and right panels.}\label{fig:lognorm}
\end{figure*}
Polarization number counts have to be assessed to know how many sources can be detected at a certain polarized flux density, $P=\sqrt{Q^2 + U^2}$, with $Q$ and $U$ being the linear polarization \emph{Stokes} parameters. Polarization measurements at mm-wavelengths are scarce because of the faintness of the polarized signal, so that both high sensitivity and robust estimates of systematic effects are required. Furthermore, completeness is very hard to be achieved with polarized samples. This is the reason why, to date, extrapolations from low frequency observations ($1.4\div 5$ GHz) are commonly adopted though the uncertainties due to intra-beam effects and bandwidth depolarization may seriously affect the estimation.      

To encompass this issue, several works in the literature \citep{2011MNRAS.413..132B,Tucci2012,Massardi2013,Bonavera2017} considered the probability function $\mathcal{P}(\Pi)$ of the polarization fraction, $\Pi=P/S$. Differential polarization number counts can be defined as
\begin{align}
n(P) = & N \int _{S_0=P} ^{\infty }   \mathcal{P}(P, S) dS = N \int _{S_0=P} ^{\infty }   \mathcal{P}(\Pi, S) \frac{dS}{S} \nonumber \\
=& \int _{S_0=P} ^{\infty }   \mathcal{P}(\Pi) n(S) \frac{dS}{S},
\label{eq:polcounts}
\end{align}
where $N$  is the total number of sources with $S \geq S_0$, $\mathcal{P}(P, S)$ and $\mathcal{P}(\Pi, S) $ are the probability functions of finding a source with flux $S$ and polarized flux $P$ or polarization fraction $\Pi$ and both can be constrained from observations.

\gpedit{ Notice that, in the last equation of \eqref{eq:polcounts}, we assume that $\Pi$ and $S$ are statistically independent. On one hand, recent results at low frequencies indicate that this  might not be the case: \citet{Stil2014} found that fainter sources ($\sim $ 1 mJy) of NVSS catalogue present a higher median fractional polarization. These results were confirmed by \citet{Lamee2016} with S-PASS: they found indications of a possible correlation between the polarization fraction and total intensity of steep-spectrum  sources  ranging from $0.42$ to $10$ Jy, whereas the correlation disappears when FSRQs are involved.  On the other hand, at higher frequencies (above $20$ GHz), \citet{Massardi2008} and \citet{Tucci2012} did not find a clear correlation between $\Pi$ and $S$ (at fluxes above $500$ mJy) for both FSRQs and SSRQs, but they found fractional polarization correlating at frequencies between  $4.8$ and $20$ GHz.}

\noindent \gpedit{ To date, surveys  at high-frequencies have not been sensitive enough to probe fainter polarized fluxes in order to seek whether this  assumption holds or not. \citet{2004MNRAS.349.1267T} further argued that at higher frequencies we observe two possible effects: (i) depolarization from Faraday rotation  is essentially negligible at frequencies above $\nu \gtrsim 10$ GHz, (ii) by  observing compact objects (\ie\ FSRQs) at increasing frequency, we probe inner and inner regions, closer to the nucleus where the magnetic field is expected to be highly ordered. Consequently if this is the case, \emph{ the polarization fraction may increase with the frequency}}. 

\noindent \gpedit{Given the goals of our work and the fact that frequencies above $10$ GHz are involved in the forecast analysis, we assume polarized fraction and flux-density uncorrelated and statistically independent but we look for some eventual dependence of $\Pi$ as function of frequency. }
  
Following \citet{2011MNRAS.413..132B}, we model $\mathcal{P}(\Pi)$ by means of a log-normal distribution,  \ie

\begin{equation}
\mathcal{P} \left(\Pi \right) = \frac{A}{\sqrt{2\pi \sigma^2} \Pi} \exp\left[  - \frac{\left( \ln(\Pi /  \mu)\right)^2}{2\sigma^2}\right]\label{eq:lognorm}
\end{equation}
where $\mu$ and $\sigma$  are respectively the median and the standard deviation in log. Notice that eq. \eqref{eq:lognorm} holds only if $0 \leq \Pi <\infty$. Although an infinite value of $\Pi$ does not have any physical meaning (synchrotron emission can be polarized up to  $ 75\%$), the values of $\mu$ and $\sigma$ are orders of magnitude smaller. Thus $\Pi$ can be effectively assumed to range up to a large value. This allows us to write a good approximation of the fractional polarization by a combination of the log-normal parameters\footnote{For further details refer to \citet{2011MNRAS.413..132B}.} 
\begin{eqnarray}
\langle \Pi  \rangle & \approx & \mu e^{\frac{1}{2} \sigma^2}, \label{eq:<pi>} \\
\langle \Pi^2  \rangle& \approx &  \mu^2 e^{2 \sigma^2}, \label{eq:<pi2>} \\
\Pi_{med} & \approx & \mu.  \label{eq:pimed} 
\end{eqnarray}

We derive polarization fraction distribution by using a \emph{bootstrap-resampling} method outlined in \citet{Austermann2009}. This generates $N_{resamp}$  simulations of the catalogue and values for unpolarized and polarized flux densities are randomly assigned for each source, from a normal distribution $\mathcal{N}(\mu_{src},\sigma_{src})$ peaking at the observed value $\mu_{src}$ and with a width $\sigma_{src}$ equal to the flux uncertainty. In the case  of upper limits, a random number is extracted from a normal distribution  centred on 0 and with width $\sigma_{src}$.  For each resampling we compute the polarization fraction and the values are distributed across bins (ranging from 5 to 15 bins depending on the number of data collected in each catalogue). The final distribution is thus  given by the mean value  within each bin and  vertical error bars computed by means of Poisson statistics, at $68\%$ of confidence level \citep[CL, ][]{Gehrels1986}, counting the observed sources in each polarization fraction bin. Finally, a log-normal distribution function \eqref{eq:lognorm} is fitted from each dataset and  $\langle \Pi \rangle$,  $\langle \Pi^2 \rangle $ and $\Pi_{med}$ are then estimated from the log-normal parameters $\mu$ and $\sigma $ as in \eqref{eq:<pi>},\eqref{eq:<pi2>},\eqref{eq:pimed}. 
\begin{table*}
\centering
\caption{Values of log-normal parameters obtained by fitting data from each catalogue. }\label{tab:bestfitlognorm}
	\begin{tabular}{l c ccc ccc l} 
\hline 
\multicolumn{9}{c}{Flat-spectrum sources}\\	
 & &&& &&& &   \\	
 $\nu $[GHz]&$N_{src}$& $A$ & $\mu$ & $\sigma$ &$\langle \Pi \rangle$& $\Pi_{med}$ & $\langle \Pi^2 \rangle^{1/2} $ &Reference \\
 \hline 
 1.4 &	82 & $0.54 \pm 0.08$ & $1.73 \pm 0.24 $ &$1.05\pm 0.09$  &$2.98 \pm 0.64 $ & $1.72 \pm 0.24$ &$5.15 \pm 1.53$ & \citet{Lamee2016}\\
  2.3 &	82 & $0.53 \pm 0.07$& $1.51 \pm 0.13 $ &$1.05\pm 0.07$  &$2.64 \pm 0.36 $ & $1.52 \pm 0.13$ &$4.59 \pm 0.87$ & \citet{Lamee2016} \\
 4.8 &	2335& $1.57 \pm 0.07$& $2.36 \pm 0.02 $ &$0.75\pm 0.01$  &$3.14 \pm 0.03 $ & $2.37 \pm 0.02$ &$4.16 \pm 0.08$ &\citet{Murphy2010}\\
 8.6 & 	2335& $1.55 \pm 0.02$& $2.46 \pm 0.01 $ &$0.73\pm 0.01$ & $3.21 \pm 0.03 $ & $2.46 \pm 0.01$ &$4.20 \pm 0.06$ &\citet{Murphy2010}\\
 8.6& 2827& $0.52 \pm 0.01$ & $2.41 \pm 0.05 $& $0.76 \pm  0.01$& $3.23 \pm 0.08 $ & $2.41 \pm 0.05$ &$4.31 \pm 0.14$ &\citet{Pelgrims2015} \\
  4.8&	109& $0.60 \pm 0.06$ & $2.02 \pm 0.13$ & $0.84 \pm 0.05$ & $2.89 \pm 0.28 $ & $2.02 \pm 0.13$ &$4.12 \pm 0.55$  &\citet{Sajina2011} \\
 8.6 &	109& $0.74 \pm 0.14$ & $2.12 \pm 0.24$ & $0.84 \pm 0.09$ & $3.01 \pm 0.51 $ & $2.12 \pm 0.23$ &$4.27 \pm 1.02$  &\citet{Sajina2011}\\
22 &	155& $1.36 \pm 0.09$ & $3.1 \pm 0.10$ & $0.88 \pm 0.03$ & $4.57 \pm 0.22 $ & $3.10 \pm 0.09$ &$6.74 \pm 0.46$ &\citet{Sajina2011}\\
43 &	111& $2.59 \pm 0.08$ & $4.48 \pm 0.06$ & $1.00 \pm 0.03$ &$7.42 \pm 0.17 $ & $4.47 \pm 0.06$ &$12.32 \pm 0.41$ &\citet{Sajina2011}\\
20 & 104 & $0.73 \pm 0.05$ & $1.73\pm 0.16$ & $0.98 \pm 0.06$ &$2.91 \pm 0.42 $ & $1.73 \pm 0.16$ &$4.89 \pm 0.99$ & \citet{Galluzzi2017b} \\ 
89 &  145 &$1.20 \pm 0.06$ &$2.86 \pm  0.10$ & $0.64 \pm 0.03$ &$3.52 \pm 0.17 $ & $2.86 \pm 0.10$ &$4.32 \pm 0.28$ & \citet{Agudo2010}\\
95 & 32 & $1.09 \pm 0.21$ & $2.13 \pm 0.23$ & $0.97 \pm 0.09$ & $3.20 \pm 0.60 $ & $2.07 \pm 0.24$ &$4.94 \pm 1.32$ & This work\\
30 & 114 & $1.51\pm 0.23$ & $2.05 \pm 0.36$ & $1.08 \pm 0.08$ & $3.69 \pm 0.92 $ & $2.06 \pm 0.37$ &$6.61 \pm 2.19$ &\citet{PlanckPCCS2} \\
44 & 30 & $2.63 \pm 0.26$ & $2.72 \pm 0.26$ & $0.77 \pm 0.11$ &  $3.69 \pm 0.66 $ & $2.73 \pm 0.26$ &$5.00 \pm 1.32$ & \citet{PlanckPCCS2}\\
70 & 34 & $3.91 \pm 0.55$ & $2.52 \pm 0.05$ & $0.58 \pm 0.06$ & $2.97 \pm 0.15 $ & $2.51 \pm 0.05$ &$3.52 \pm 0.30$ & \citet{PlanckPCCS2} \\
100& 20 & $2.18 \pm 0.28$ & $5.15 \pm 0.69$ & $0.80\pm 0.10$ &$7.19 \pm 1.59 $ & $5.17 \pm 0.73$ &$9.99 \pm 3.07$  &\citet{PlanckPCCS2} \\
143& 25 & $3.13\pm 0.10$ & $5.98 \pm 0.16$ & $0.80 \pm 0.04$ &$8.39 \pm 0.35 $ & $6.02 \pm 0.13$ &$11.69 \pm 0.80$ & \citet{PlanckPCCS2}\\
217& 11 & $3.44 \pm 0.32$ & $3.74 \pm 0.29$ & $0.88 \pm 0.11$ & $5.47 \pm 0.34 $ & $3.70 \pm 0.27$ &$8.09 \pm 1.09$ &
\citet{PlanckPCCS2}\\
\hline
\multicolumn{9}{c}{Steep-spectrum sources}\\
 & &&& &&& &   \\	
1.4 & 388& $1.12 \pm 0.08$& $1.47 \pm 0.11$ & $1.05\pm 0.08$ &$2.56 \pm 0.35 $ & $1.47 \pm 0.11$ &$4.45 \pm 0.95$ & \citet{Lamee2016}\\
2.3& 388  & $1.78 \pm 0.07$& $1.93 \pm 0.06 $ &$0.80\pm 0.05$&$2.66 \pm 0.14 $ & $1.93 \pm 0.06$ &$3.66 \pm 0.31$ &\citet{Lamee2016} \\
4.8&952 & $2.07 \pm 0.07$ & $2.83 \pm 0.07 $ &$0.81\pm 0.04$ &$3.92 \pm 0.15 $ & $2.84 \pm 0.08$ &$5.43 \pm 0.37$ & \citet{Murphy2010}\\
8.4& 952& $3.02 \pm 0.03$ & $2.13 \pm 0.12 $ &$1.13\pm 0.03$ &$4.85 \pm 0.05 $ & $3.02 \pm 0.05$ &$7.79 \pm 0.18$ & \citet{Murphy2010}\\
20 & 952 & $4.55 \pm 0.12$ & $6.98\pm 0.12 $ & $0.55 \pm 0.01$ & $8.10 \pm 0.18 $ & $6.98 \pm 0.12$ &$9.41 \pm 0.27$ &\citet{Murphy2010} \\ 
4.8&	39& $2.72 \pm 0.65$ & $2.35 \pm 0.46$ & $1.07 \pm 0.43$ & $4.19 \pm 1.42 $ & $2.35 \pm 0.42$ &$7.49 \pm 5.61$ &\citet{Sajina2011}\\
8.6 &39& $1.94 \pm 0.14$  & $3.39 \pm 0.31$ & $1.04 \pm 0.10$ &$5.90 \pm 1.10 $ & $3.41 \pm 0.32$ &$10.23 \pm 2.95$ &\citet{Sajina2011} \\
22 &	 38 & $2.51 \pm 0.10$ & $5.76 \pm 0.19$ & $0.82 \pm 0.05$ &  $8.08 \pm 0.44 $ & $5.73 \pm 0.17$ &$11.40 \pm 0.99$ &\citet{Sajina2011}\\
43 &	 15& $4.74  \pm 0.08$& $9.89 \pm 0.20$ & $0.73 \pm 0.02$ &$12.43 \pm 0.29 $ & $9.62 \pm 0.13$ &$16.06 \pm 0.55$ & \citet{Sajina2011}\\
\hline 
\end{tabular}
\end{table*}
We show in \figref\ \ref{fig:lognorm} the polarization fraction distributions from PACO-ATCA at 20 GHz and PACO-ALMA at 95 GHz (the best fit parameters of the other datasets used in this analysis are summarized in \tabref\ \ref{tab:bestfitlognorm}). 
We show in top panel of \figref\ \ref{fig:polcounts} the polarization number counts computed by \citet{Galluzzi2017b} at $20$ GHz (blue circles) as a result of the convolution of total intensity number counts with the log-normal distribution $\mathcal{P}(\Pi )$ as in eq. \eqref{eq:polcounts}. We further overlap the predicted  total counts from both the D05 (solid thin blue) and C2Ex (solid thin gray) models convolved with the distribution function.  As already stated in \secref\ \ref{sec:numbcounts}, at $20$ GHz both models are equivalent even for polarized number counts. 

\gpedit{ In bottom panel of \figref\ \ref{fig:polcounts} are shown the polarized number counts  at 95 GHz coming from the PACO-ALMA  sample of 32 sources as lower green triangles. Given the paucity of this sample, we re-sample it by means of 1,000 bootstrap-resampling. The resampled source counts (shown as green lower triangles in \figref\ \ref{fig:polcounts})  are then computed in a similar manner as for the $20$  GHz observations and are summarized in the companion paper by Galluzzi et al. (2018, in prep.). The error bar estimation  of each data point  include the Poissonian $68 \% $ CL uncertainties  \citep{Gehrels1986} plus  the error  derived from the uncertainties of log-normal parameters $\delta_A,\delta_{\mu}, \delta_{\sigma}$ (summarized in \tabref\ \ref{tab:bestfitlognorm}). This error has been assessed by means of  differencing the  number counts convolved with an upper and a lower log-normal function, respectively estimated at maximum and minimum values of log-normal parameters. }
  
 We would like to stress that this is the first time that number counts from the PACO-ALMA sample have been computed and exploited for this kind of analysis. Notice that the data are very well fitted by both predictions.
 
\begin{figure*}[htb]
\includegraphics[width=1\columnwidth]{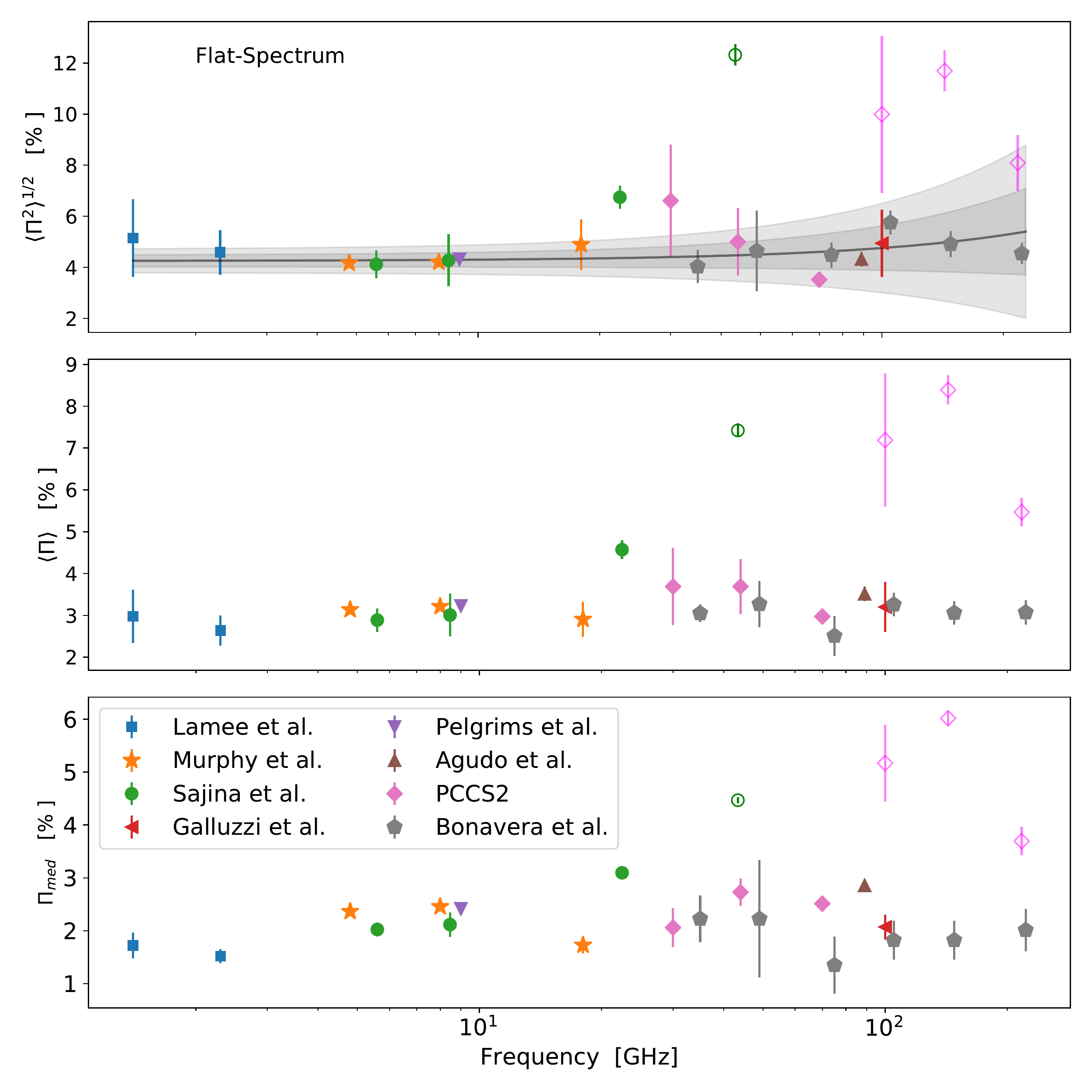}
\includegraphics[width=1\columnwidth]{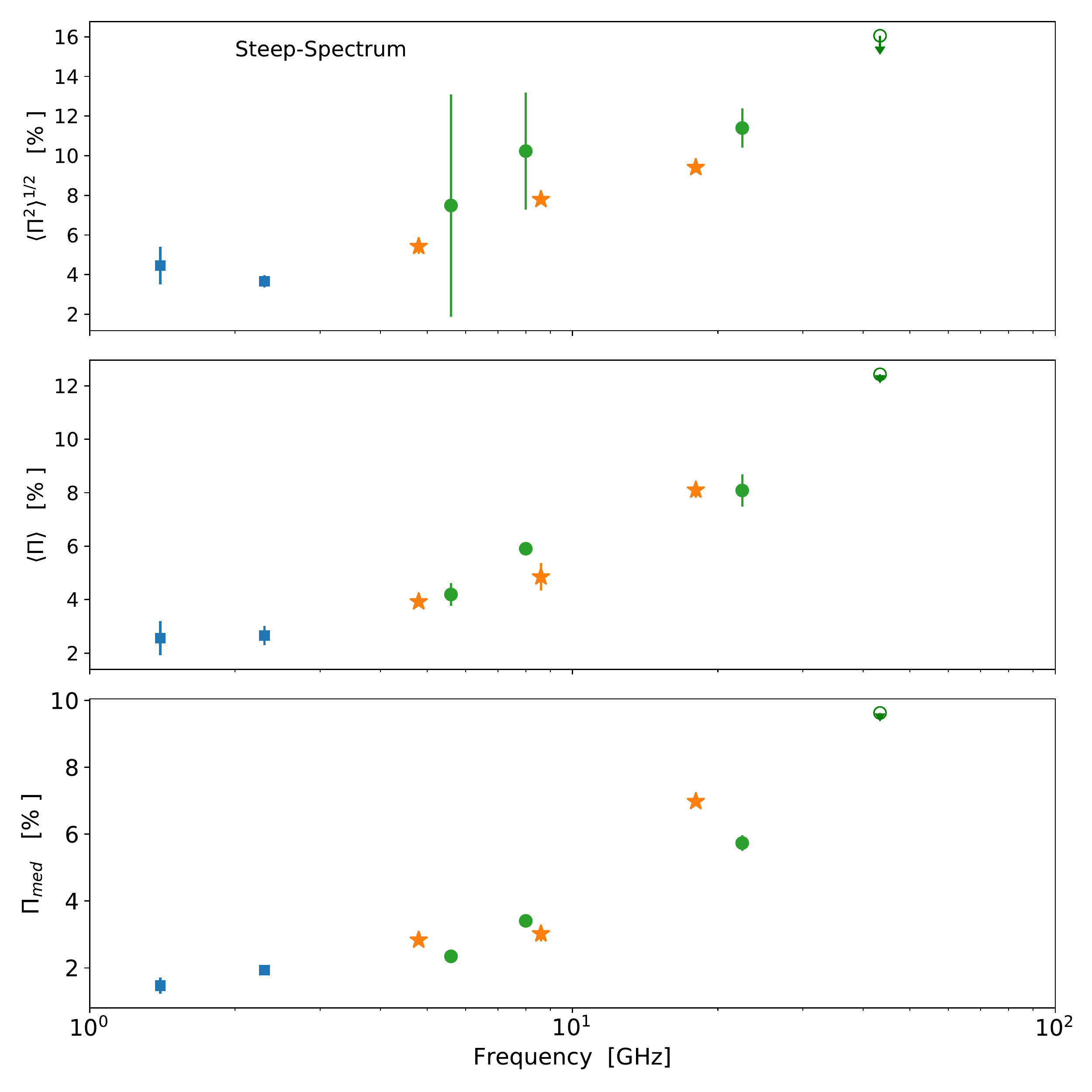}
\caption{ Values of ${\langle \Pi^2 \rangle}^{1/2}$ (top), $\langle \Pi \rangle$ (center)  and $\Pi_{med} $ (bottom) derived from best fit lognormal parameters \eqref{eq:<pi2>}, \eqref{eq:<pi>}, \eqref{eq:pimed}. Open symbols are data which have not been taken into account for the linear fit in \eqref{eq:fitpi}. We distinguished FSRQs (left) from the SSRQs (right). In top left panel, a linear function is fitted from the data to provide a scaling of polarization fraction as a function of frequency. Light and dark shaded area are respectively  the $1\sigma$  and $2\sigma$ uncertainties.}\label{fig:pi2}
\end{figure*}

The estimated values of $\langle \Pi \rangle$, $\Pi_{med}$ and $\langle \Pi^2 \rangle ^{1/2} $ for FSRQ(left panel) and SSRQ (right panel) are shown in  \figref\ \ref{fig:pi2}. \gpedit{By comparing the two panels, we note that the SSRQ fractional polarizations increases with frequency.  Although this could be simply related to observational bias (at higher frequencies, steep-spectrum sources contributes at fainter fluxes), such frequency dependence of $\Pi$ for SSRQs has been already discussed in \citet{Tucci2012}. On the contrary, the fractional polarization measured for the FSRQ remains almost constant during the frequency range studied.}
\gpedit{To quantify this  dependence, we estimate a linear fit on $\langle \Pi^2 \rangle ^{1/2} $  as a function  of  a wide (around 2 orders of magnitude) range of frequencies. This choice is mainly due to the fact that $\langle \Pi^2 \rangle $ values  are needed to estimate B-mode angular power spectrum of polarized ERSs and we include in the linear fit also the values of $\langle \Pi^2 \rangle ^{1/2} $ estimated by \citet{Bonavera2017} between 30 and 217 GHz. They were derived assuming a log-normal distribution as in this work. 
 In particular, for the  best fit, we retain only fractional polarization from the FSRQs and BL Lacs since their contribution dominates number counts at larger fluxes and at frequencies $>20 $ GHz (see \figrefs\ \ref{fig:polcounts} and  \ref{fig:counts}). The linear fit involves the data for which the estimation of $\mu $ and $\sigma$ are reliable (filled symbols in \figref\ \ref{fig:pi2}). Open symbols indicate data that have not been included to the fit, mainly because of the poor statistics in fitting the log-normal distribution (\eg\ less than 20 polarized sources have been detected in polarization in the \pla\ HFI channels, see \tabref\ref{tab:bestfitlognorm})}.
 
  We find a negligible frequency dependence of  $\langle \Pi^2 \rangle ^{1/2}$: 
\begin{eqnarray}
\langle \Pi^2 \rangle^{1/2}(\nu )  &=  \left( {0.005 } \pm 0.006 \mathrm{ GHz^{-1}}\right) \nu  \nonumber \\&+  \left({4.170}\pm 0.22 \right).\label{eq:fitpi}
\end{eqnarray}
In the top left panel of \figref\ \ref{fig:pi2} we show the linear fit as a gray solid line with darker and lighter shaded areas resembling  respectively the $1\sigma$ and $2 \sigma $ uncertainties on best fit parameters. Notice that for $\nu >20 $ GHz, we found $\langle \Pi^2 \rangle ^{1/2} \sim 4 \% $, in agreement with the value found by  \citet{Tucci2012} and consistent with the expectations of \citet{2004MNRAS.349.1267T}  and \citet{Stil2014}.

At l $\nu < 20 $ GHz, SSRQs have to be taken into account to forecast the contribution of ERS to \cmb\ observations.  Thus,  we  perform the same linear fit  by including SSRQs for all the datasets at frequencies smaller than $20 $ GHz, shown in \figref\  \ref{fig:pi2} (top right panel). The best fit  equation  changes into  
\begin{eqnarray*}
\langle \Pi^2 \rangle^{1/2}(\nu )  &=  \left(-0.015 \pm 0.009 ) \mathrm{\,GHz}^{-1}\right) \nu  \nonumber \\&+  \left(5 .43 \pm  0.23 \right).\label{eq:2fitpi}
\end{eqnarray*}
Nonetheless the slope is still negligible,  the presence of SSRQs enhances the average polarization fraction of sources at  frequencies $\nu \lesssim 20 $ GHz and, as one can notice in \figref\ \ref{fig:pi2}, this is consistently observed in ${\langle \Pi\rangle}$ as well. 

\gpedit{We would like to stress that  selection effects could bias our  results towards larger values of $\Pi$, especially where few tens of polarized sources have been detected, see \tabref\ \ref{tab:bestfitlognorm}. This is the reason why we excluded PCCS2 HFI data (magenta diamonds) in \figref\ \ref{fig:pi2} and we considered the ones from \citet{Bonavera2017} (gray pentagons). To this regard, the stacking technique helps because it includes the faint sources to the statistical estimate of $\Pi$  even if those sources are not directly detectable. }

\section{Forecasts for forthcoming \cmb\  ground-based experiment}\label{sec:4casts}

In this section we present the forecast analysis for current and forthcoming \cmb\ surveys performed with a \texttt{Python} package Point Source ForeCast (PS4C) made publicly available\footnote{\url{https://gitlab.com/giuse.puglisi/PS4C}}.  PS4C is a user friendly platform which allows to forecast the contribution of radio point sources both in total intensity and polarized flux-densities  given the nominal specifics of a \cmb\ experiment.
In \tabref \ref{tab:specs} we summarize the specifics of 5 \cmb\ experiments with whom we forecast  the  ERS contribution with PS4C:
 \begin{itemize}
  \item the Q-U-I JOint TEnerife  \citet[QUIJOTE][]{2014arXiv1401.4690L} CMB experiment designed to observe the polarized emissions from the \cmb, our Galaxy and the  extra-galactic sources at four frequencies in the range between  10 and 20 GHz  and at FWHM resolution of $\sim1^{\circ}$. Observations started observing in  November 2012, covering $18,000\, \mathrm{deg}^2$  of the Northern hemisphere, and achieved the sensitivity of $1800{\,\mu}$K arcmin in polarization;
 \item a generic  \cmb-S2 experiment observing at $95,\, 150$ GHz within a patch including $ 2 \%$ of the sky at the resolution of $3.5 $ arcmin, at $25\div 30\, \mu\mathrm{K\,arcmin}$ sensitivity;
 \item a \cmb-S3 ground based experiment with the   so-called  \emph{strawman} configuration, as it has been defined in \citet{2016arXiv161002743A}, for the ``measuring-r'' survey. It  consists of an array of small-aperture (SA, $\sim 1$ m) telescopes and one  large-aperture (LA, $\sim 5$ m) telescope, observing at the accessible atmospheric windows in the sub-millimeter  range  (at about  30, 40, 90, 150 GHz). The sensitivities at these frequencies are targeted to be about  $1\div10 \,\mu $K arcmin.
 \item the Lite satellite for the studies of B-mode polarization and Inflation from cosmic Background Radiation Detection \citep[LiteBIRD][]{Matsumura2016} is a   satellite mission proposed to JAXA  aimed at  measuring the \cmb\ polarized signal at degree angular scale. Its goal  is to characterize the  measurement  of   $r$ with an uncertainty $\sigma( r )<0.001$. In order to achieve such high accuracy, the target detector sensitivity is  $2\, \mu$K arcmin observing over a wide range of frequencies (from  $40$ to $320$ GHz). The current effort aims to launch in 2025;
\item \gpedit{ the Cosmic ORigin Explorer \citep[CORE]{2017arXiv170604516D} is a next generation space-borne experiment and it has been proposed as a Medium-size ESA mission opportunity. It has been designed as the \pla\ satellite successor, planned to have better angular resolution  and sensitivity than \pla . We consider the  \emph{CORE150} configuration: a satellite involving a $1.5$ m telescope, observing  over a wide range of frequency channels (up to $800 $ GHz) with sensitivities ranging from $\sim 10 $ to $5 \, \mu\mathrm{K\, arcmin}$. In this work, we restrict our analysis to a selection of frequency channels, (see the last row of \tabref\ \ref{tab:specs}) to compare the expectations with  the ones  previously obtained by \citet{2016arXiv160907263D}.}
 \end{itemize}

\begin{table*}[htb]
\centering
\caption{Nominal specifics of  \cmb\ experiments described  in \secref\ref{sec:4casts}. }\label{tab:specs}
	\begin{tabular}{lcccc} 
		\hline
 & Frequency [GHz] & Sensitivity  $\left[ \mu\mathrm{ K \, arcmin}\right]$  & FWHM & $f_{sky}$\\
\hline
QUIJOTE & 11,13,17,19 & 1800 & $1^{\circ} $&  $50\%$ \\
\cmb-S2 & 95, 150 & 25,30 & $3.5^{\prime}$  & $5 \%$ \\
{\cmb-S3} SA& 30, 40, 95,150 & $8,\, 6,\,1,\, 2\, $& $1^{\circ} $& $20 \%$\\
\cmb-S3 LA & 30, 40, 95,150& $8,\, 6,\,1,\, 2\,$ & $10',\, 7',\,3',\, 2' $ &$20 \%$ \\
\multirow{2}{*}{LiteBIRD} & 40,   50,   60,   68,   78 & $53,32,25,19,15  $  & $1^{\circ} $&  $100\%$ \\
& 89, 100,119, 140,166 & $ 12,  15.6,12.6,8.3,8.7 $& $1^{\circ} $&  $100\%$ \\
CORE150 &  $60,100,145 $ &  $10.6, 7.1,5.1 $ & $14', 8', 6'$ & $100 \%$\\ 
\hline
\end{tabular}
\end{table*}

\begin{table*}[htb]
\centering
\caption{ Number of polarized  ERS detected above the $P_{3\sigma}$ flux density detection limit in polarization,  by current and forthcoming \cmb\ ground based experiments. Counts are  estimated both from the D05  and  the C2Ex predictions (in brackets).}\label{tab:cmbground}
\begin{tabular}{l cc| cc cc} 
\hline
&		\multicolumn{2}{c}{\cmb\ -S2} &\multicolumn{4}{c}{\cmb\ -S3 }\\
			&				 &		    &{    SA}& & {  LA}& \\
$\nu $[GHz] & $P_{3\sigma} $ [mJy]& $N_{3\sigma}$& $P_{3\sigma}$ [mJy]& $N_{3\sigma}$ & $P_{3\sigma}$ [mJy]& $N_{3\sigma}$\\
\hline
{30} & \dots&	\dots&15 &236 (191) &1.5&2329 (2278)\\
{40}&  \dots	&\dots	& 15&215	(156)&1.5 &1867 (1810) \\
{95}& 100	& 3 (2)& 10 &355 (222)& 1&2432 (2136)\\ 
{150}& 100&3 (1)&15 & 146 (74)& 1.5&1145 (867)\\  
\hline
\end{tabular}
\end{table*}
\begin{table}[h]
\centering
\caption{ Number of sources detected above the $\geq S_{\lim} $  and $\geq P_{lim}$ flux densities limit by the QUIJOTE experiment, assuming the nominal and conservative values for sensitivity. Values are estimated using D05 and C2Ex models (ins brackets). }\label{tab:quijote}
\begin{tabular}{l cc cc} 
\hline
$\nu $[GHz] & $S_{lim} $ [Jy]& $N_{src}$&$P_{lim} $ [Jy]& $N_{src}$\\
\hline
\multirow{2}{*}{11} &0.5 &   694 (673)     & 0.5& 6 (4)\\
				   &  1 & 347 (340) & 1 &  2 (1)	\\
\multirow{2}{*}{13}&  0.5& 445 (434)&  0.5 &2 (1)\\
				  &	1	& 210 (205) &	1 &0 (0)	\\
\multirow{2}{*}{17}& 1	&201 (197)	&	1	  & 0 (0)\\ 
				  &	2	&86		(83) &	2     &	0 (0)\\
\multirow{2}{*}{19}& 1 	&	128 (125)	&	1	&0 (0)\\  
				   &	2 	&52 	(51) 	&	 2 	& 0  (0)  \\
\hline
\end{tabular}
\end{table}

 Although most of the  frequency channels of future experiments range up to $350$ GHz, we forecast up to $150$ GHz. This is because  at higher frequencies the contribution coming from dusty galaxies and Cosmic Infrared Background cannot be neglected\footnote{ We  have  already planned to include into the package the contribution from dusty galaxies and forecasts with PS4C  will be presented in a future release that will be described in a future paper.} \citep{2013MNRAS.429.1309N,2016arXiv160907263D}.  \citet{2017MNRAS.472..628B} estimated the polarized contribution of dusty galaxies  by stacking about 4700 sources observed by \pla\ at  $143, \,217, \,353$ GHz HFI channels. They estimated the polarized contribution of dusty galaxies to B-mode power spectra and found that at frequencies larger than $217$ GHz these population of sources might remarkably contaminate the primordial B-modes.
 
\begin{table}[htpb]
\centering
\caption{Number of sources observed above $3 \sigma_{det} $ limit in terms of polarized flux density $P_{3\sigma}$ by the LiteBIRD experiment. Bracketed values are estimated using the C2Ex model. }\label{tab:litebird}
\begin{tabular}{l  cc } 
\hline
$\nu $ [GHz]& $P_{3\sigma} $ [mJy]& $N_{3\sigma}$\\
\hline
40& 	450&4 (3)\\
50&	240& 11 (8)\\	
60&	210 &9 (6)\\
68&	300& 4 (3)\\
78&	 240&	6 (4)\\
89&	210& 12 (8)\\
100&240& 10 (7)\\
119& 210&	14 (10)\\
140&  270&	8 (4)\\
166&270&  7 (4)\\
\hline
\end{tabular}
\end{table}

 We compute one realization of \cmb\  power spectra by means of the CAMB  package \citep{Lewis:1999bs} by assuming the \pla\ best fit cosmological parameters \citep{2016A&A...594A..13P} and a tensor to scalar ratio $r=0.05$ (slightly below the current upper limits).

 To assess the contribution of ERS to the power spectrum level, we assume their distribution in the sky to be Poissonian, since the contribution of clustering starts to be relevant for  $S<10$ mJy \citep{2005ApJ...621....1G,2005A&A...438..475T}. The power spectrum of temperature fluctuations coming from a Poissonian distribution of sources is expected to be a constant contribution at all multipoles. \gpedit{In particular, we consider as \emph{masked} all sources whose flux-density is above $3\sigma$ the detection limit $S_{cut}= 3 \sigma_{det} $ and we do not include them into power spectrum estimate}
\begin{equation}
C_{\ell} ^{T} = \left( \frac{dB}{dT} \right) ^{-2} N \langle  S^2\rangle=\left( \frac{dB}{dT} \right) ^{-2} \int_0 ^{S_{cut}} n(S) S^2 dS, 
\label{eq:clt}
\end{equation}
where $n(S)$ and $N$ are respectively the differential and the integral number counts per steradian, and $dB/ dT$ is the conversion factor from brightness to temperature, being 
\begin{displaymath}
\left( \frac{dB}{dT} \right)^{-1} \approx 10^{-2}  \frac{(e^x -1)^2}{x^4 e^x} \frac{\mu \mathrm{K} }{\mathrm{Jy \, sr^{-1}}} , 
\end{displaymath}
 with $x = \nu / 57 $ GHz. \citet{2004MNRAS.349.1267T} found that it possible to relate the ERS polarization power spectrum to the intensity one  \eqref{eq:clt} as follows  
 \begin{eqnarray}
 C_{\ell} ^Q= & \left( \frac{dB}{dT} \right) ^{-2} N \langle Q^2 \rangle \nonumber \\
 =& \left( \frac{dB}{dT} \right) ^{-2} N \langle  S^2 \Pi^2 \cos^2 2 \phi \rangle \nonumber \\ 
 =& \left( \frac{dB}{dT} \right) ^{-2} N \langle  S^2\rangle \langle \Pi^2 \rangle \langle \cos^2 2 \phi \rangle \nonumber \\ 
 =&\frac{1}{2} \left( \frac{dB}{dT} \right) ^{-2}  \langle \Pi^2 \rangle C_{\ell }^T, 
\label{eq:clp}
 \end{eqnarray}
where the $1/2$ factor comes from the average value of $\cos^2 2 \phi $, if the polarization angle $\phi$ is uniformly distributed. The value for $\langle \Pi^2 \rangle $ is derived at each frequency from eq.\eqref{eq:fitpi}.  Since we do expect point sources to  equally contribute on average both to $Q$ and $U$, and thus to the $E$-  and $B$- modes, we can approximate $ C_{\ell} ^B\simeq  C_{\ell} ^E\simeq C_{\ell} ^U\simeq C_{\ell} ^Q$. In the following, B-mode  power spectra are normalized by the usual normalization factor $\mathcal{D}_{\ell} =\ell (\ell +1)C_{\ell}/2\pi$. 

\gpedit{To forecast the number of sources that will be detected in intensity and polarized flux-density above a given  detection limit, we integrate the differential number counts, $n(S)$ and $n(P)$ as}
\begin{align}
N(>S)=& \int_{S_{cut}} ^{\infty}  n(S)  dS, \label{eq:Ns}\\
N(>P)=& \int_{P_{cut}} ^{\infty}  n(P)  dP.\label{eq:Np}
\end{align}

Finally, \gpedit{to compare the level of contamination produced by the ERS with the Galactic foreground one,} we rescale the Galactic foreground emission at a given $f_{sky}$, frequency $\nu$ and multipole order $\ell$ as in \citet{pldiffuse2015},
{\small
\begin{align}
\mathcal{D}^{FG}(\ell, \nu , f_{sky } ) =&   \frac{Var  \left[ \mathrm{Sync},f_{sky}\right]}{Var  \left[\mathrm{Sync},f_{sky,0}\right]} q_{s }  \left( \frac{\ell}{80} \right) ^{\alpha_{s}} \frac{s_{s}(\nu)}{s_{s} (\nu_s)}  + \nonumber \\
& \frac{Var \left[ \mathrm{Dust},f_{sky}\right]}{Var \left[\mathrm{Dust},f_{sky,0}\right]} q_{d }   \left( \frac{\ell}{80}\right) ^{\alpha_{d}} \frac{s_{d}(\nu)}{s_{d} (\nu_d)} .
\label{eq:clgal}
\end{align}}
with $s, d$ referring respectively to synchrotron and dust.\gpedit{ For all the   parameters  entering in \eqref{eq:clgal}, we use the best fit  values  quoted in  \citet[table 11][]{pldiffuse2015} estimated outside the Galactic plane in the  \texttt{UPB77}  mask \citep[][defined in section 4.2]{2016A&A...594A...9P}. The mask has been computed considering a common foreground mask after component separation analysis with $1^{\circ}$ apodization scale. } 
Therefore, to rescale the estimate  in eq.\eqref{eq:clgal}to  a patch with a smaller fraction of sky, $f_{sky}$,  we need to  compute the variance of both synchrotron and thermal dust template maps within the considered  patch and  within the \pla\ region with $f_{sky,0}= 73\%$. The rescaled foreground power spectra are shown in \figref\ \ref{fig:powerspectra} as dotted lines.

\subsection{PS4C with current and forthcoming \cmb\ ground based experiments}

\begin{figure*}[h]
\includegraphics[width=1\columnwidth]{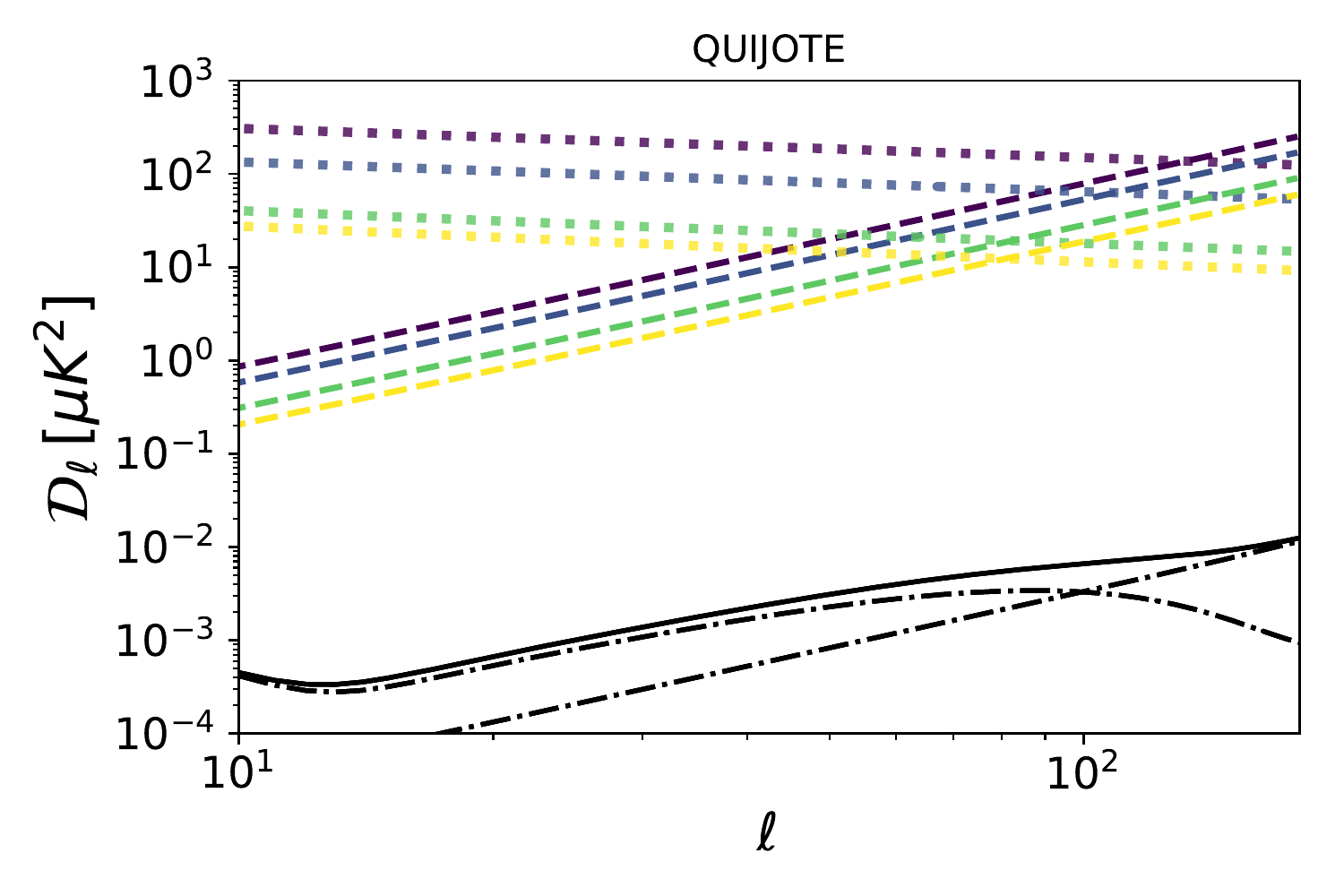}
\includegraphics[width=1\columnwidth]{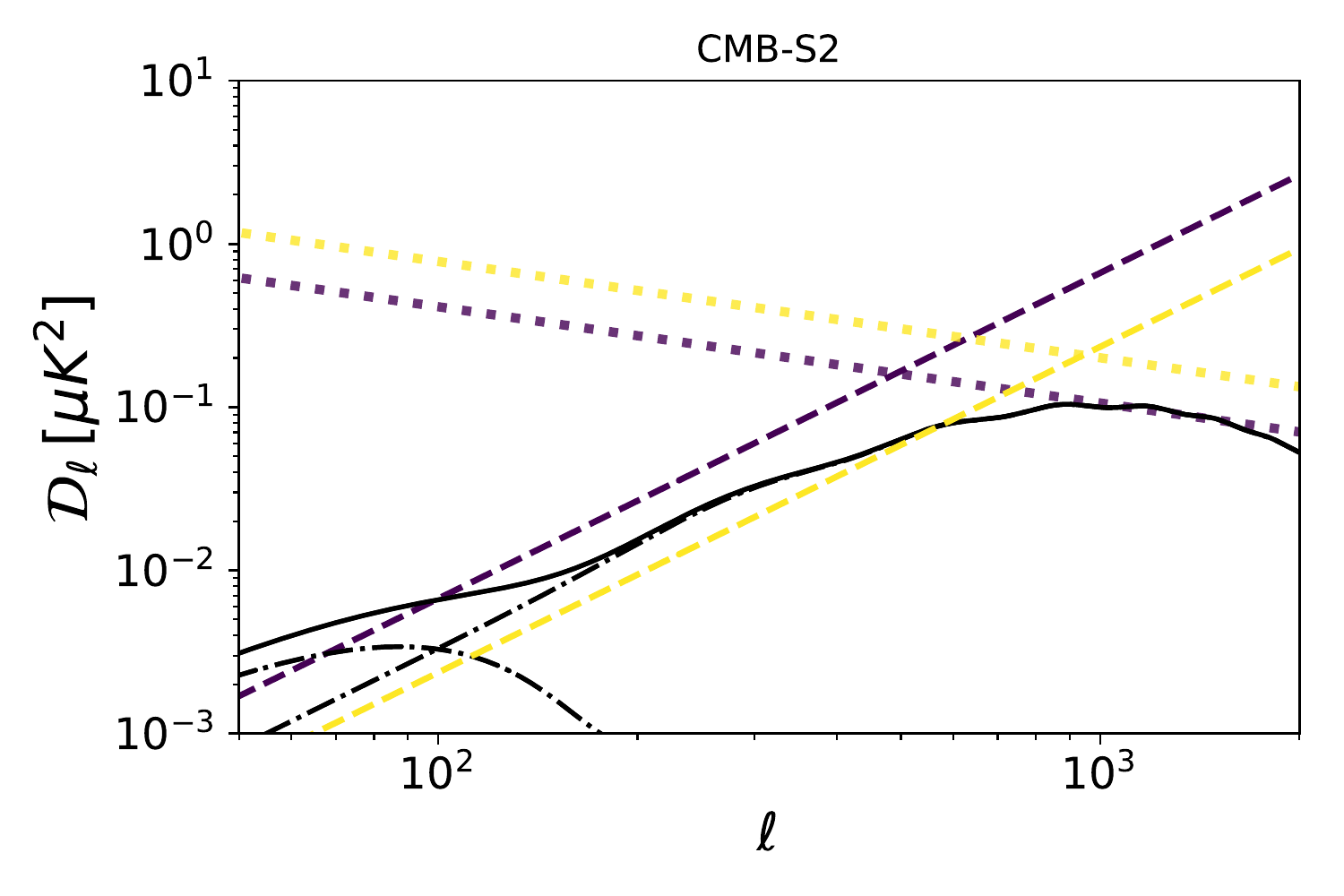}\\
\includegraphics[width=1\columnwidth]{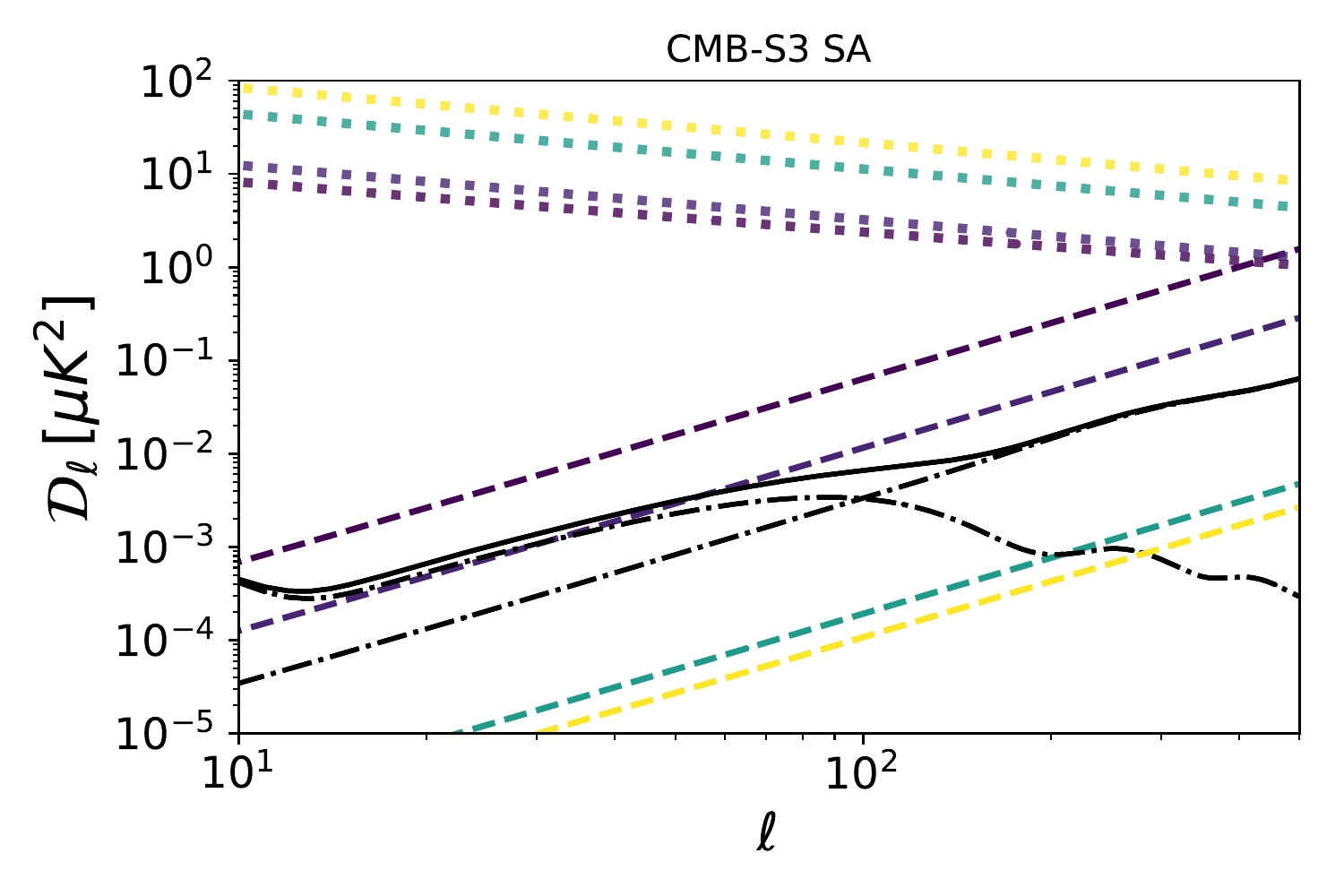}
\includegraphics[width=1\columnwidth]{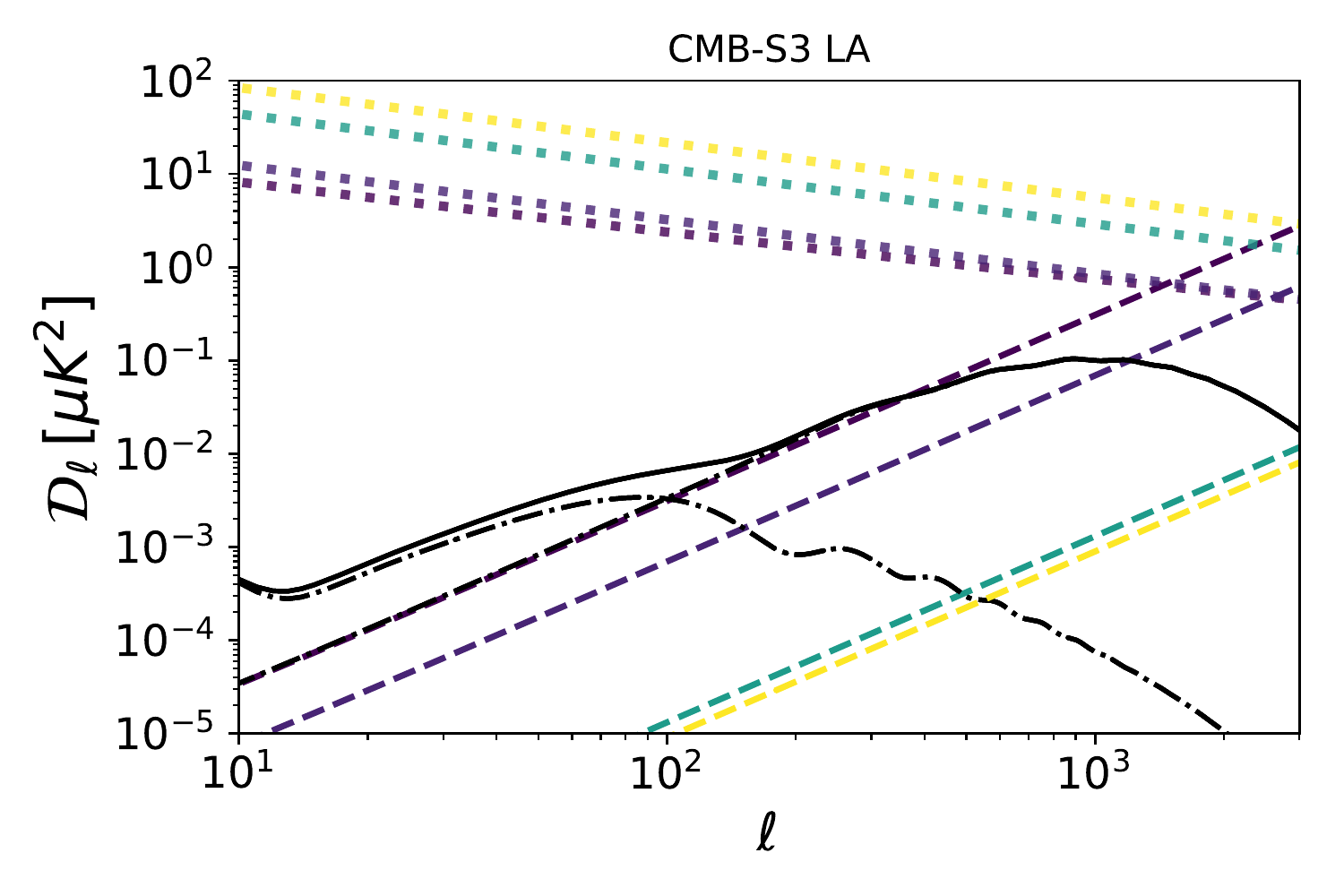} \\
\includegraphics[width=1\columnwidth]{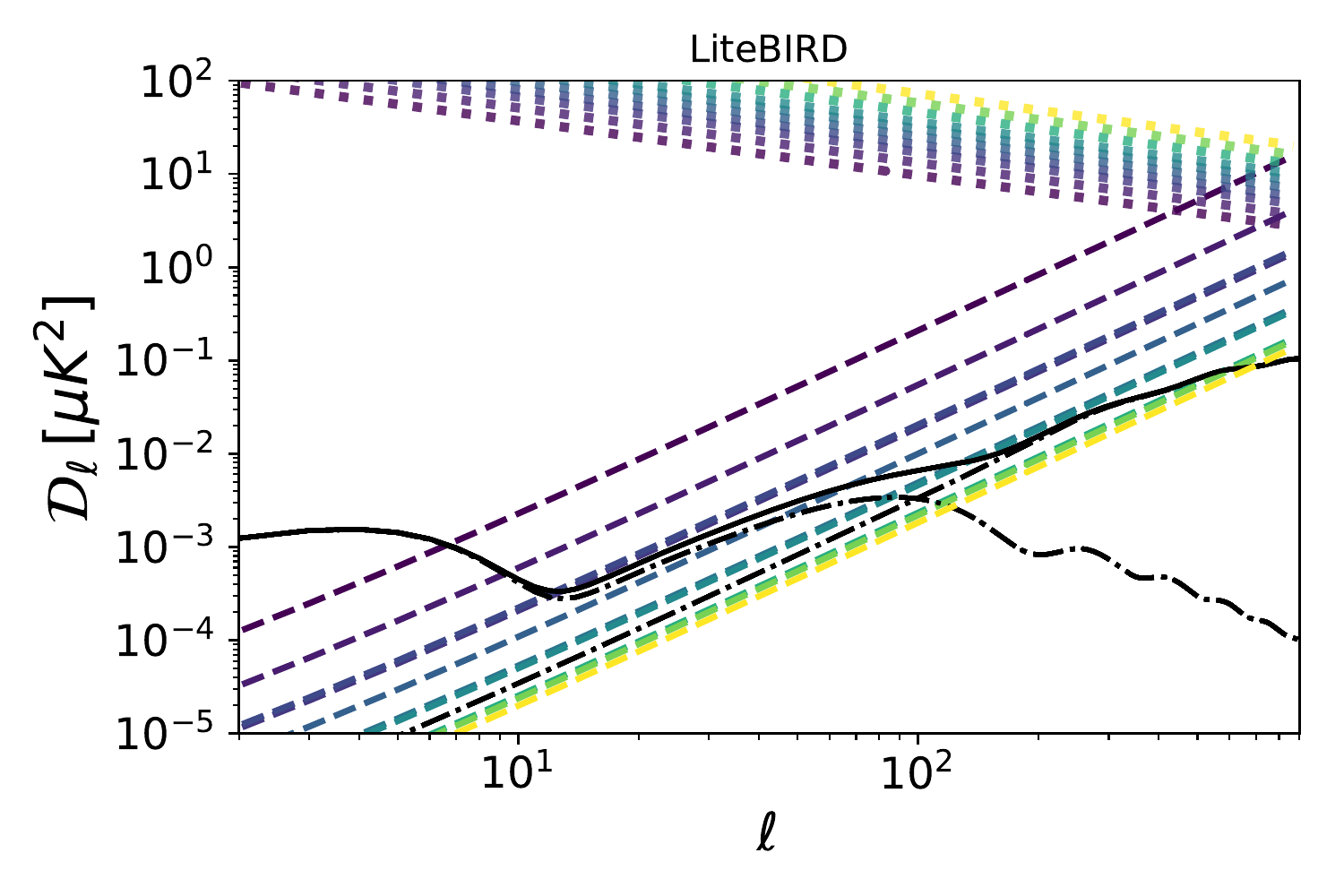}
\includegraphics[width=1\columnwidth]{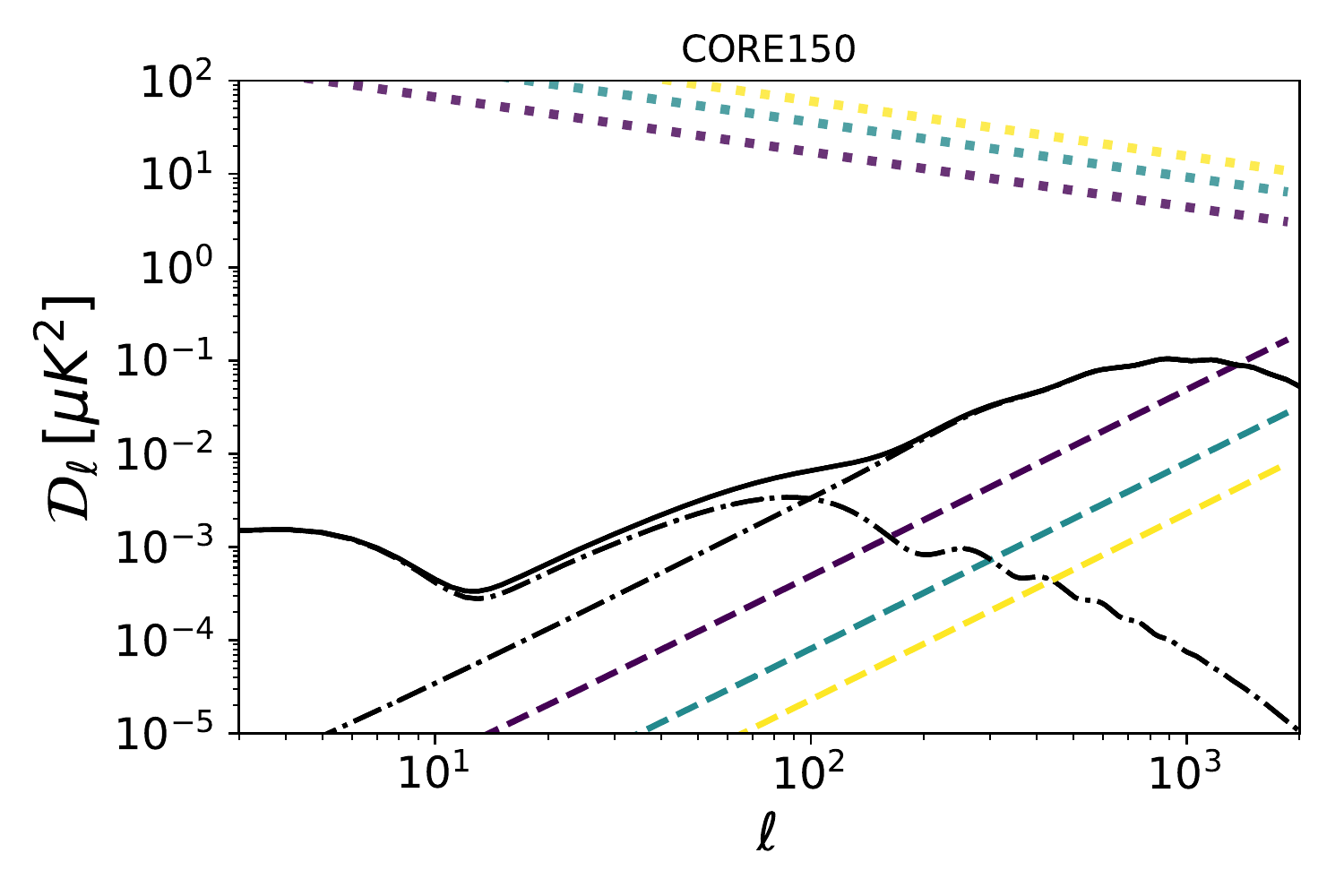}
  \caption{Forecasts of foreground contamination with PS4C. In all panels, the black dot-dashed lines show the primordial ($r=0.05$) and lensed \cmb\ B-mode power spectra and the black solid line is the the total CMB B-mode power spectrum. \gpedit{The dotted (dashed) lines are the power spectrum of the polarized Galactic emission (ERS emission) at the different frequencies available for each experiment, the color scale is such that the colors go from purple to yellow as the frequency increases. The power spectra depend are estimated using eq.\eqref{eq:clgal} in the region outside the \texttt{UPB77 } \pla\ mask (in order to exclude the Galactic plane and ERS above the $3\sigma $ detection limit). The different panels corresponds to predictions for different experiments. From top to bottom and from left to right: QUJOTE ($11$, $13$, $17$, $19$ GHz), \cmb-S2 ($95$ and $150$ GHz), \cmb-S3 observing with small and large aperture telescopes ($30$, $40$, $95$, $150$ GHz), LiteBIRD (frequencies between $40-166$ GHz) and CORE150 ($60,\, 100,\,145$ GHz.}}\label{fig:powerspectra}
\end{figure*}

\figref\ \ref{fig:powerspectra} shows our PS4C forecasts of foreground contamination to the recovery of the CMB B-mode for the different experiments in the different panels: we plot the expected spectrum in polarization of Galactic (dotted lines) and ERS (dashed lines) emissions at the different frequencies available for each experiment and the total CMB B-mode power spectrum (black solid line). The black dot-dashed lines show the primordial ($r=0.05$) and lensed B-mode power spectra separately. The power spectra are computed in the region outside the \texttt{UPB77 } \pla\ mask (in order to exclude the Galactic plane and the ERS whose flux density is below the $3\sigma $ detection limit). 
The Galactic foreground turns out to be the most contaminating emission in the B-mode recovery. The different colors for the Galactic and ERS spectra are for different frequencies, going from purple to yellow as the frequency increases.
It should be commented that there exists several \emph{component separation} and \emph{foreground cleaning} algorithms that can recover \cmb\ intensity and polarization signals with great accuracy \citep{pldiffuse2015}. In addition, multi-frequency observations and joint analyses from different experiments \citep{PhysRevLett.114.101301} can improve the foreground cleaning. So, even if in our work we are considering the most conservative cases, it should be stressed that such contamination could be lowered \citep[ at sub-percentage level][]{2016PhRvD..94h3526S,2011PhRvD..84f9907E} by applying such foreground removal algorithms. 

\gpedit{In particular, \figref\ \ref{fig:powerspectra} shows our forecasts for the QUIJOTE (top left) and \cmb-S2 (top right) experiments. 
As for QUIJOTE, the Galactic emission is much higher than the \cmb\ one and higher than the contribution from undetected ERS, except at small angular scales where the ERS start to be dominant. Since the QUIJOTE experiment ranges from $10$ to $20$ GHz, we need to take into account the contribution from both FSRQs and SSRQs}, with the resulting increase in the average fractional polarization and number counts (see \figref\ \ref{fig:pi2} and \figref\ \ref{fig:counts}). \tabref\ \ref{tab:quijote} summarizes the total number of sources in total intensity (third column) and polarization (fourth column) that QUIJOTE would detect (frequencies are given in the first column), assuming nominal and conservative sensitivity values (flux density limits in total intensity and polarization are listed in columns two and three respectively). We found $694$, $445$, $201$ and $128$ sources in total intensity at $11$, $13$, $17$, $19$ GHz respectively. In polarization only a few of them would be detected and just in the $11$ and $13$ GHz channels.

\gpedit{ For the \cmb-S2 experiment whose frequencies are greater than 95 GHz, the Galactic emission (mostly thermal dust emission) is the most contaminating up to $\ell  \sim 350$, while the ERS are important at small angular scales. Unlike the previous case, at these frequencies the \cmb\ B-mode spectrum is comparable to the one of undetected ERS.}
     
\begin{figure*}[htpb]
\includegraphics[width=1\columnwidth]{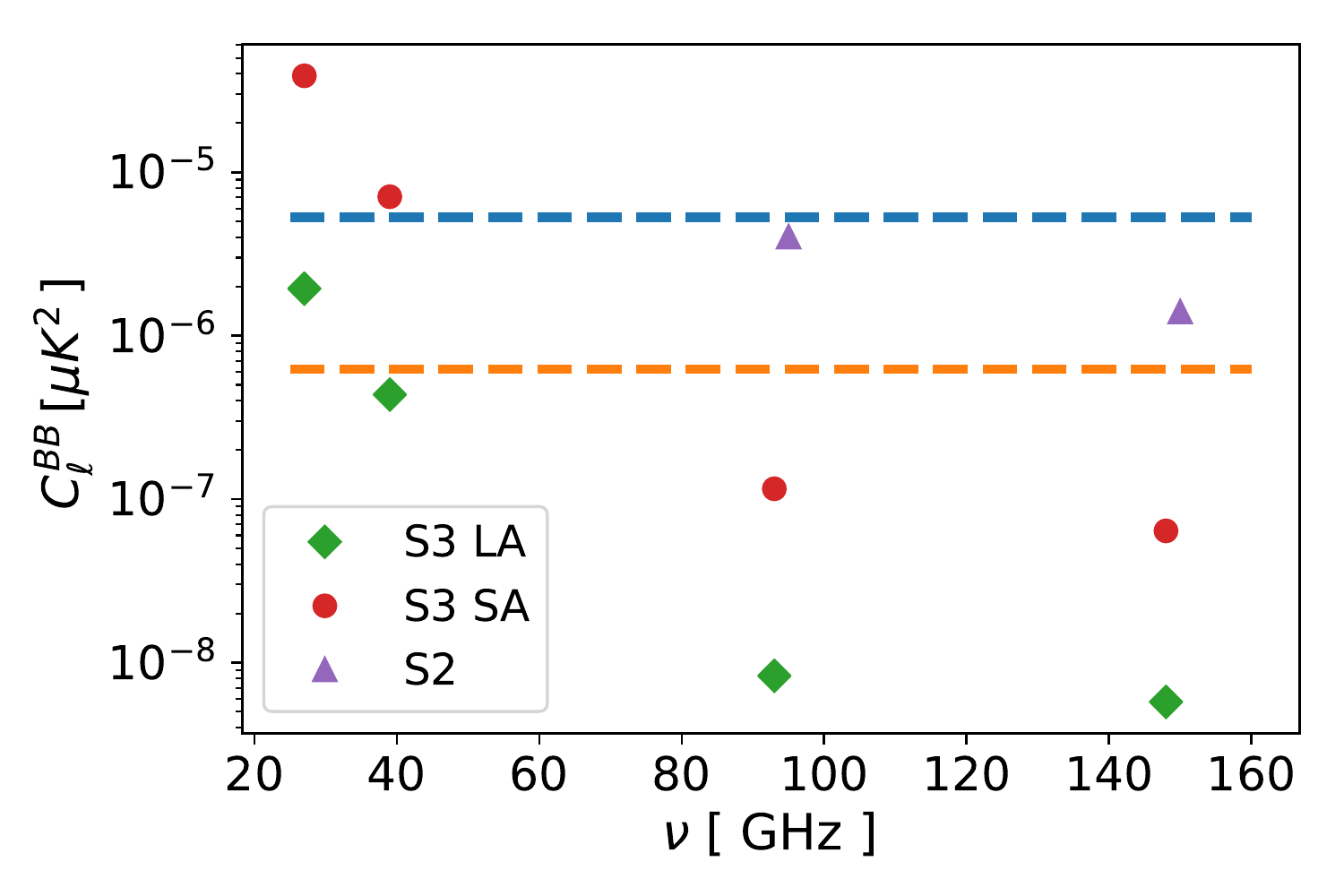}
\includegraphics[width=1\columnwidth]{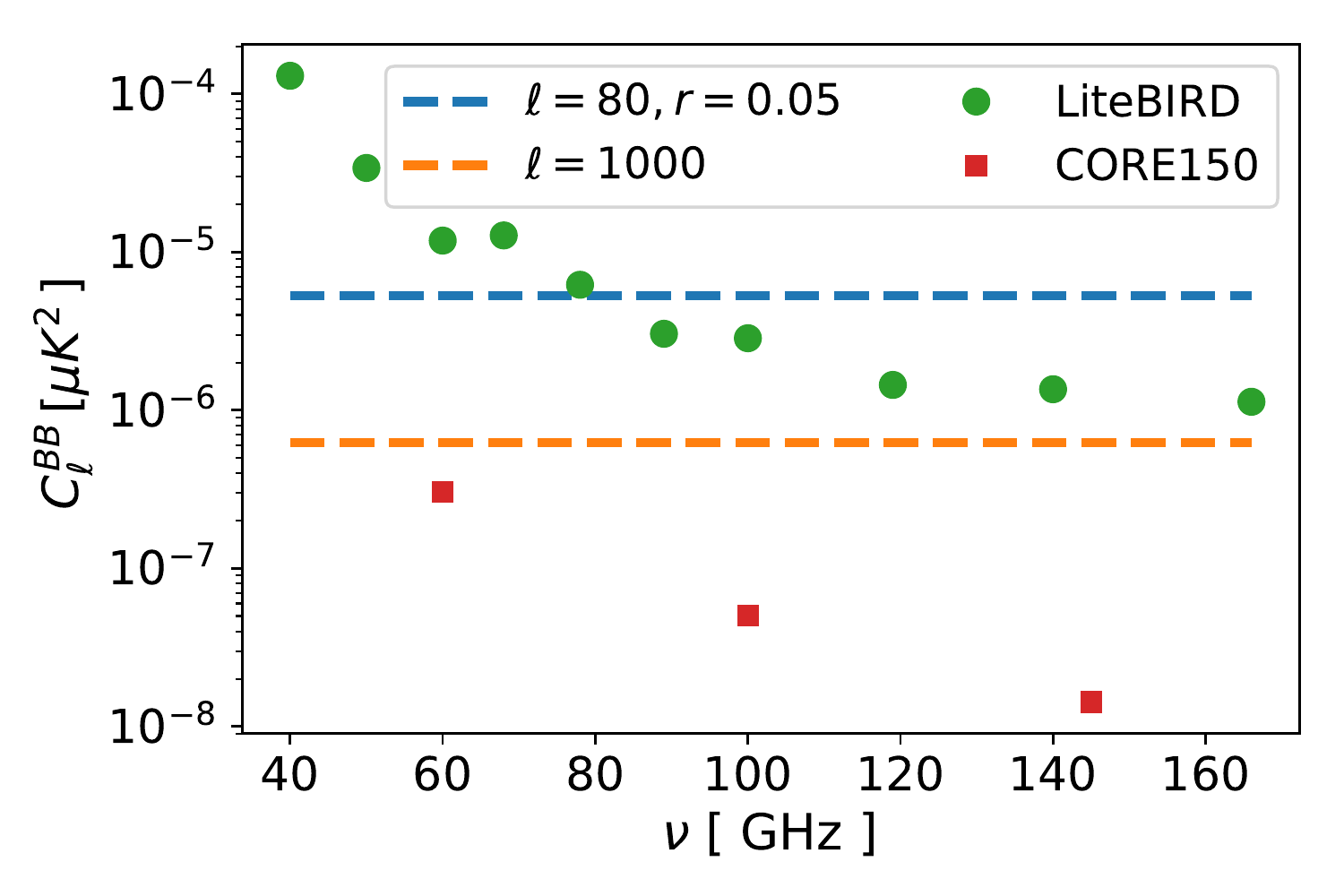}
\caption{Power spectra in polarization of undetected ERS in current and future \cmb\ experiments. Left panel: \cmb-S2 (triangles) and \cmb-S3  (circles for the small aperture telescope and diamonds for the large aperture telescope). Right panel: LiteBIRD (circles) and CORE150 (squares). The dotted lines are the B-mode power spectra at the acoustic scale ($\ell=80$) and at the lensing B-modes peak scale ($\ell\approx 1000$).} \label{fig:fc_w_freqs}
 \end{figure*}

 \gpedit{ \noindent In \figref\ \ref{fig:fc_w_freqs} the triangles show the $C_{\ell}^{BB}$ of undetected ERS estimated using eq.\eqref{eq:clp}. The detection limits are given by the \cmb-S2 sensitivities. The $C_{\ell}^{BB}$ of the CMB B-mode are also plotted: the cyan dashed line is for the case $\ell\approx 80$ and $r=0.05$ and the orange dashed line is for $\ell\approx 1000$. \figref\ \ref{fig:fc_w_freqs} shows what is the contamination due to undetected ERS and consequently the level of source detection required to  detect primordial or lensing $B$-mode signal. In \cmb-S2 the undetected ERS level of the power spectrum is comparable to the lensing B-mode one. In this case, given the experiment sensitivity and the size of the observed region, $\sim 150$ sources would be detected in total intensity and only few of them in polarization at a $3\sigma $ level.}

Among the experiments studied in this work, the \cmb-S3 is the one with the greatest sensitivity and best resolution. The results are shown in the central panels of \figref\ \ref{fig:powerspectra} and in the left panel of \figref\ \ref{fig:fc_w_freqs} with circles and diamonds. As summarized in \tabref\ \ref{tab:cmbground}, the maximum number of polarized sources detected above a $3\sigma$ level and using the large aperture telescope is $2329$ with flux density $P_{lim} \gtrsim 1 $ mJy. When using a smaller aperture telescope, this number drops to a few hundreds with polarized flux densities $P_{lim} \gtrsim 10 $ mJy.

The contribution in polarization of undetected ERS is very small at high frequencies($\nu \gtrsim 90$) and at low multipoles $\ell \lesssim 2000$.  At lower frequencies, undetected ERS still can contaminate and they have to be taken into account 
to de-lens, lensing $B$-modes to get the primordial ones for $r \lesssim 0.05$. 

\subsection{PS4C with future space missions}

The results for the LiteBIRD experiment are shown in the left bottom panel of \figref\ \ref{fig:powerspectra} and the filled circles in the right panel of \figref\ \ref{fig:fc_w_freqs}. On the whole, the most contaminating contribution is the Galactic one, except at small angular scales ($l \sim 400$) and high frequencies ($\nu > 70$ GHz) where the ERS contribution is comparable to the Galactic one. The ERS contribution, although generally lower than the Galactic one, is also important being higher than the CMB B-mode level even at large scales ($l \gtrsim 7$) and $\nu < 70$ GHz (dashed purple and blue lines). Moreover, at $\nu > 80$ GHz and $l \gtrsim 70$ the ERS contribution is comparable to the B-mode power spectrum. The number of sources that would be detected in polarization above the $3\sigma$ level with this experiment are listed in \tabref\ \ref{tab:litebird} and they range from 4 at 10 and 68 GHz to 14 at 119 GHz. The first column is the frequency in GHz, the second is the polarized flux density limit in mJy and the third column is the number of sources that would be detected by LiteBIRD (values in the brackets are estimated from the C2Ex model).

\gpedit{Our findings for {CORE} are shown in the right bottom panel of \figrefs\ \ref{fig:powerspectra} and in the right panel of \figref\ \ref{fig:fc_w_freqs} (squares). Galactic emission is the most contaminating for B-mode detection. Undetected  ERS are important only at 60 GHz, where their power spectrum is comparable to the one of the B-mode due to lensing. {CORE} would be able to detect up to 200 sources per steradian, implying a lower contamination for the CMB B-mode power spectrum with respect to LiteBIRD.}

\gpedit{\tabref\ \ref{tab:comparison} compares the surface densities (i.e. number of sources per steradian, last two columns) at {CORE} frequencies (first column) of the polarized ERS above the $P_{4 \sigma} $ flux density limit (second column) estimated by \citet{2016arXiv160907263D}(DZ16) and PS4C (values in the brackets are for C3Ex estimate). In this comparison we use a $4 \sigma $ flux density limit in order to be consistent with the estimates by \citet{2016arXiv160907263D}. 
Above $100$ GHz, we find a discrepancy between D05 and DZ16 that could be due to two effects that become more important at higher frequencies: (i) the D05 predictions tend to over-estimate the polarized source number counts (see \secref\ \ref{sec:numbcounts}) and (ii) at $\nu>100$ the polarization fraction is expected to suffer a slight increase (from $\sim 4\%$ to $\sim 5\%$ from $100$ to $150$ GHz) as can be seen in eq. \eqref{eq:fitpi} and \figref\ \ref{fig:pi2}. }

\gpedit{On one hand, at $100$ GHz, we find that accounting solely for the  observation in (ii), i.e. a $20 \%$ increase of $\Pi $ to a value of $4.67\%$, the D05 forecasts predict source counts  that are $20 \%$ larger than DZ16\footnote{ For this estimate, we assume  that differential source counts are described by a power law with  spectral index $ >1$}. On the other hand, at $150$ GHz, the surface density estimated with PS4C with D05 model is $\sim 65\%$ larger than the value referred by DZ16.  Even accounting for the $25\%$ fractional increase of $\Pi $   to  $4.92\%$ from eq.\eqref{eq:fitpi}, this is not enough to compensate the observed discrepancy. We thus argue that the discrepancy at  $150$ GHz is caused by  both (i) and (ii). }

\gpedit{Contrary to the D05 forecasts, the C2Ex model is in reasonable agreement with \citet{2016arXiv160907263D}, meaning that the C2Ex predictions are more robust than the D05 ones at least at higher frequencies.}

\begin{table}[htpb]
\centering
\caption{ Comparison of surface densities  of polarized ERSs brighter than $P_{4 \sigma} $  estimated by \citet{2016arXiv160907263D}(DZ16) and by PS4C. Values in brackets refer to C2Ex  estimates. }\label{tab:comparison}
\begin{tabular}{l  cc c } 
\hline
$\nu $ [GHz]& $P_{4\sigma } $ [mJy]& \multicolumn{2}{c}{ $N_{4\sigma}\,  [ \, \mathrm{sr^{-1}} \, ] $}  \\
 & & DZ16& PS4C \\ 
\hline
60&	5.2  &212 &214  (198)\\
100& 5.2	 &184& 229  (164)\\
145&  4.6&165& 271	(142)\\
\hline
\end{tabular}
\end{table}

\section{Summary and conclusions} \label{sec:conclusions}

We describe and present the state-of-the-art observations on polarization of  ERS over a wide  frequency range, namely from $1.4$ to $ 217$ GHz. We exploit for the first time the polarization number counts  at 95 GHz from  a sample of 32 polarized sources detected with  ALMA. The characterization of these sources and their spectral behaviour in frequencies ranging from  $1 $ to $95$ GHz are described in a companion paper by Galluzzi et al. (2018, in prep.)

By collecting polarization flux densities from 10 catalogues, we are able to derive a relation of the average fractional polarization  as a function of frequency and to avoid extrapolations that have been commonly adopted to forecast the average polarization fraction  from low- ($\lesssim 20$ GHz  where enough data have been collected), to high-frequency ($\gtrsim 70$ GHz where still few polarization measurements have been performed).
 Therefore, we fit a linear function on data from several surveys, including \pla\ measurements from both detection and stacking, and we find a mild dependence of $\langle \Pi ^2 \rangle ^{1/2}$ as a function of $\nu$. 

This relation allows us to forecasts the contribution of ERSs to polarization B-mode power spectrum given the nominal sensitivities of current  and forthcoming \cmb\ experiments, by means of predictions of ERS counts coming from two models D05 and C2Ex.  The whole forecast suite is fully integrated into  a \texttt{Python}  package,  PS4C, made publicly available with online  documentation and tutorials.

We discuss the reasons why we do not assume a correlation between the level of fractional polarization and the total intensity flux. Although still controversial and not observed at high-radio frequencies \citep[Galluzzi et al. 2018 in prep.,][]{Galluzzi2017b,Galluzzi2017a,Massardi2013}, deeper surveys in polarization are critical to further proof the validity of this assumption, not only at higher frequencies but also at fainter flux density levels. 

Future \cmb\ experiments could shed light on  this interesting aspect:  in fact, we have shown that they are going to observe an increasing number of polarized ERS (they are foreseen to detect up to $\sim 2000$ polarized ERS) because their sensitivity will increasingly improve in the future. 

A further potentiality of future \cmb\ experiments is that they can be largely exploited by the community as \emph{wide global surveys} to measure  polarized flux density of sources at very high-radio frequencies \citep{2017Galax...5...47P}. Programs  aimed at observing ERSs at  higher resolution  can thus  benefit of \cmb\ large area surveys in an extremely wide range of frequencies, from $20 $ up to $300$ GHz. 

 Moreover, since  in this work we mostly focus on blazar statistical polarization, as it is the main bright source population at frequencies $< 150 $ GHz, we restrict our forecast  analysis up to this frequency limit. At higher frequencies, the far-IR dusty star forming  galaxies constitute the majority of extra-galactic sources (see \citet[fig.25][]{PlanckPCCS2}) and, similarly to the ERSs, their polarized emission contaminates  B-mode power spectra\footnote{In addition to the Poissonian contribution,  an extra-term coming from  clustering has to be considered when dusty sources are involved.} \citep{2015JCAP...06..018D}. Recent works from \citet{2017MNRAS.472..628B,2016arXiv160907263D} have already shown statistical polarization properties of dusty sources and forecasted  their contribution for future \cmb\  experiments. Therefore, we plan to  include those estimates within the PS4C package in a future development. 

As a final remark, we stress that ERSs below the detection flux limit may introduce a bias at all the angular scales and at frequencies $\nu< 50$ GHz: their synchrotron emission is still strong enough to contaminate polarization measurements even at low flux densities, namely $P< 1$ mJy. At larger frequencies, ERS polarization power spectra have to be assessed as long as smaller angular scales  are involved to estimate the \cmb\ power spectrum at multipoles around the lensing  peak or to estimate the primordial B-mode power spectrum at lower multipoles ($\ell < 800$) by means of de-lensing algorithms.

\acknowledgments
\small{\emph{Acknowledgments.} We are pleased to thank  Matteo Bonato, Bruce Partridge, Vincent Pelgrims  and the referee for their  helpful comments and suggestions. This work was supported by the RADIOFOREGROUNDS grant of the European Union's Horizon 2020 research and innovation programme (COMPET-05-2015, grant agreement number 687312) and by  by National Institute of Nuclear Physics (INFN) INDARK. LB and JGN acknowledge financial support from the I+D 2015 project AYA2015-65887-P (MINECO/FEDER). J.G.N also acknowledges financial support from the Spanish MINECO for a ‘Ramon y Cajal’ fellowship (RYC-2013-13256). VG and MM acknowledge financial support by the Italian {\it Ministero dell'Istruzione, Universit\`a e Ricerca} through the grant {\it Progetti Premiali 2012-iALMA} (CUP C52I13000140001).} We further acknowledge support from the ASI-COSMOS Network\footnote{\url{www.cosmosnet.it}}. 




\bibliographystyle{aasjournal}
\bibliography{ps}


\appendix
\renewcommand\thefigure{\thesection.\arabic{figure}}    
\section{ }
\setcounter{figure}{0}    
\begin{figure*}[hb]
\includegraphics[width=1\columnwidth, trim= 0 15cm 0 0 ,clip=true]{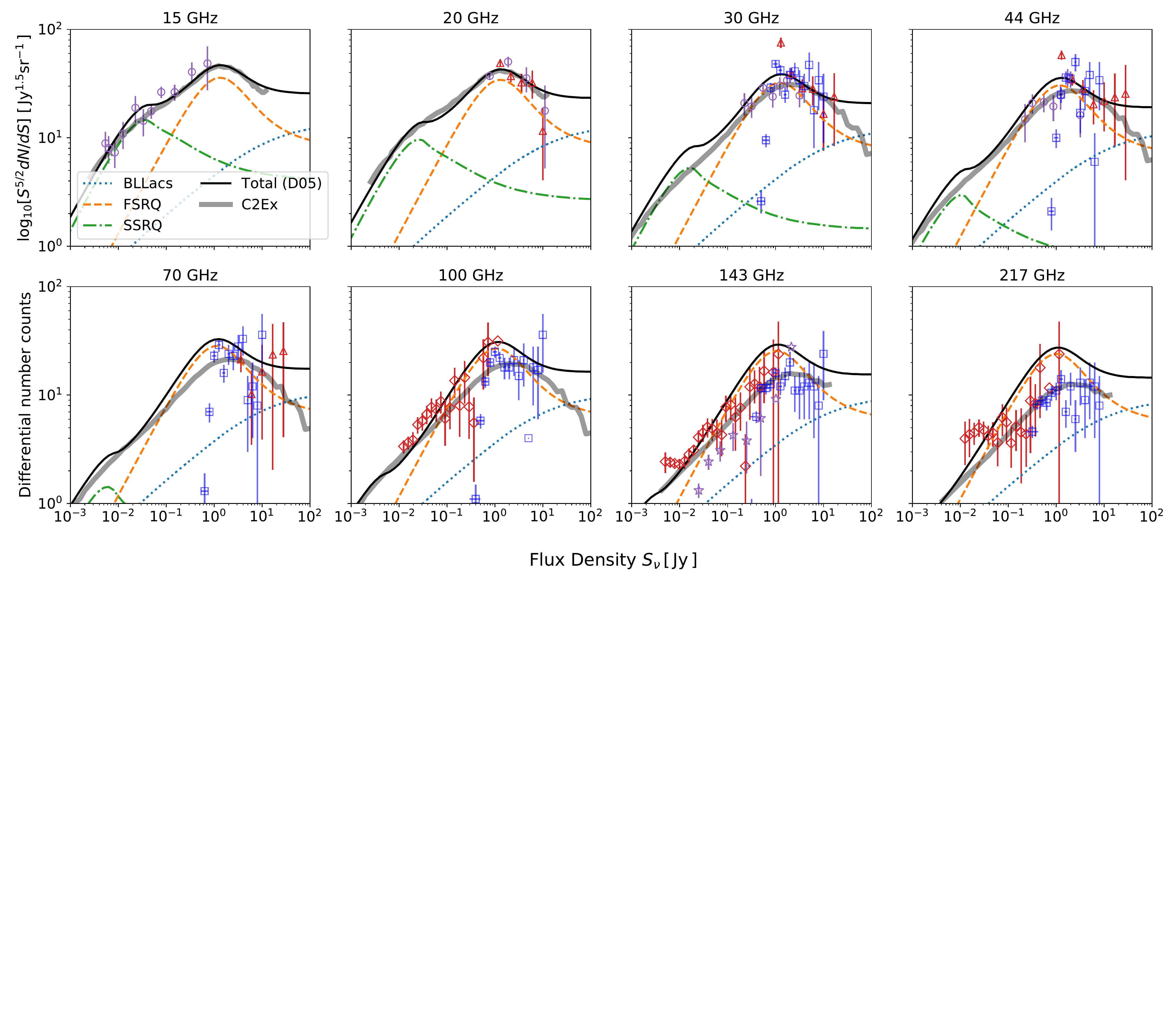}
\caption{Euclidean normalized differential number counts on a wide range of frequencies. The dotted, dashed, dot-dashed and solid lines are respectively the number counts of BL Lacs, FSRQs, SSRQs and their total contribution predicted by the D05 model. The thick solid grey line  are the number counts coming from the C2Ex model. Number counts obtained with several experiments observing at similar frequency channels are also shown: open circles at 15 and 20 GHz are respectively from \citet{2003MNRAS.342..915W,2010MNRAS.404.1005W} and \citet{Massardi2008}; open circles at 30  and 44 GHz resemble counts from the PACO dataset  \citep{2011MNRAS.416..559B}; upper triangles from 20 to 70 GHz are data from WMAP5-yr survey  \citep{2009MNRAS.392..733M}); diamonds at 100, 143, 217 GHz are from SPT  \citep{2013ApJ...779...61M}; stars at 143 GHz are ACT  counts  \citep{2011ApJ...731..100M}; squares  at 30, 44, 70, 100, 143, 217 GHz are based on data from \citet{PlanckearlyXII}.}\label{fig:counts}
\end{figure*}

\end{document}